\documentclass[twocolumn, prX, showkeys]{revtex4}
\usepackage{amsmath,amssymb,multirow,amsfonts,amsthm}
\usepackage[pdftex]{graphicx}
\usepackage[tight]{subfigure}
\usepackage[all]{xypic}
\usepackage[numbered,framed]{mcode}

\theoremstyle{remark}
\newtheorem{remark}{Remark}

\begin{document}
	
	\title{Subset Simulation Method for Rare Event Estimation: An Introduction}
	
	\author{Konstantin Zuev}
	\affiliation{Institute for Risk and Uncertainty, University of Liverpool, Liverpool, UK}
	\email{K.Zuev@liverpool.ac.uk}
	
	\begin{abstract}
		
		This paper provides a detailed introductory description of Subset
		Simulation, an advanced stochastic simulation method for estimation
		of small probabilities of rare failure events. A simple and intuitive derivation of the method is given along with the discussion on its implementation.	The method is illustrated with several easy-to-understand examples. For demonstration purposes, the MATLAB code for the considered examples is provided.
		The reader is assumed to be familiar only with elementary
		probability theory and statistics.
		
	\end{abstract}


\maketitle

\section{Introduction}

Subset Simulation (SS) is an efficient and elegant method for
simulating rare events and estimating the corresponding small tail
probabilities. The method was originally developed by Siu-Kui Au and
James Beck in the already classical paper \cite{AuBeck} for
estimation of structural reliability of complex civil engineering
systems such as tall buildings and bridges at risk from earthquakes.
The method turned out to be so powerful and general that
over the last decade, SS has been successfully applied to
reliability problems in geotechnical, aerospace, fire, and nuclear
engineering. Moreover, the idea of SS proved to be useful not only in
reliability analysis but also in other problems associated with
general engineering systems, such as sensitivity analysis, design
optimization, and uncertainty quantification. As of October 2013,
according to the Web of Knowledge database, the original SS paper
\cite{AuBeck} received more than 250 citations that
indicates the high impact of the Subset Simulation method on the
engineering research community.

Subset Simulation is essentially based on two different ideas:
conceptual and technical. The conceptual idea is to decompose the
rare event $F$ into a sequence of progressively ``less-rare'' nested
events,
\begin{equation}\label{idea1}
    F= F_m\subset F_{m-1}\subset\ldots\subset F_1,
\end{equation}
where $F_1$ is a relatively frequent event. For example, suppose
that $F$ represents the event of getting exactly $m$ heads when
flipping a fair coin $m$ times. If $m$ is large, then $F$ is a rare
event. To decompose $F$ into a sequence (\ref{idea1}), let us define
$F_k$ to be the event of getting exactly $k$ heads in the first $k$
flips, where $k=1,\ldots,m$. The smaller $k$, the less rare the
corresponding event $F_k$; and $F_1$
--- getting heads in the first flip --- is relatively frequent.

Given a sequence of subsets (\ref{idea1}), the small probability $\mathbb{P}(F)$ of
the rare event $F$ can then be represented as a product
of larger probabilities as follows:
\begin{equation}\label{idea2}
\begin{split}
  \mathbb{P}(F)&= \mathbb{P}(F_m)\\
  &=\mathbb{P}(F_{1})\frac{\mathbb{P}(F_{2})}{\mathbb{P}(F_{1})}
  \frac{\mathbb{P}(F_{3})}{\mathbb{P}(F_{2})}\ldots\frac{\mathbb{P}(F_{m-1})}{\mathbb{P}(F_{m-2})}\frac{\mathbb{P}(F_m)}{\mathbb{P}(F_{m-1})}  \\
    &=\mathbb{P}(F_1)\cdot\mathbb{P}(F_2|F_1)\cdot\ldots\cdot\mathbb{P}(F_m|F_{m-1}),
\end{split}
\end{equation}
where
$\mathbb{P}(F_{k}|F_{k-1})=\mathbb{P}(F_{k})/\mathbb{P}(F_{k-1})$
denotes the conditional probability of event $F_{k}$ given the
occurrence of event $F_{k-1}$, for $k=2,\ldots,m$. In the coin
example, $\mathbb{P}(F_1)=1/2$, all conditional probabilities
$\mathbb{P}(F_{k}|F_{k-1})=1/2$, and the probability of the rare
event $\mathbb{P}(F)=1/2^m$.

Unlike the coin example, in real applications, it is often not
obvious how to decompose the rare event into a sequence
(\ref{idea1}) and how to compute all conditional probabilities in
(\ref{idea2}). In Subset Simulation, the ``sequencing'' of the rare
event is done adaptively as the algorithm proceeds. This is achieved
by employing Markov chain Monte Carlo, an advanced simulation
technique, which constitutes the second -- technical -- idea behind
SS. Finally, all conditional probabilities are automatically
obtained as a by-product of the adaptive sequencing.

The main goals of this paper are: (a) to provide a detailed
exposition of Subset Simulation at an introductory level; (b) to
give a simple derivation of the method and discuss its
implementation; and (c) to illustrate SS with intuitive examples.
Although the scope of SS is much wider, in this paper the method is
described in the context of engineering reliability estimation, the
problem SS was originally developed for in \cite{AuBeck}.

The rest of the paper is organized as follows.
Section~\ref{Reliability Problem} describes the engineering
reliability problem and explains why this problem is computationally
challenging. Section~\ref{DMCsection} discusses how the direct Monte
Carlo method can be used for engineering reliability estimation and
why it is often inefficient. In Section~\ref{Transformation}, a necessary preprocessing step
which is often used by many reliability methods is briefly discussed.
Section~\ref{SS} is the core of the paper, where the SS method is explained.
Illustrative examples are considered in Section ~\ref{Examples}.
For demonstration purposes, the MATLAB code for the considered examples is provided in Section~\ref{Appendix}.
Section~\ref{Conclusions} concludes the paper with a brief summary.

\section{Engineering Reliability Problem}\label{Reliability Problem}

One of the most important and computationally challenging problems
in reliability engineering is to estimate the probability of failure
for a system, that is, the probability of unacceptable
system performance. The behavior of the system can be described
by a \textit{response variable} $y$, which may represent, for
example, the roof displacement or the largest interstory drift. The
response variable depends on \textit{input variables}
$x=(x_1,\ldots,x_d)$, also called \textit{basic variables}, which may
represent geometry, material properties, and loads,
\begin{equation}\label{performance function}
    y=g(x_1,\ldots,x_d),
\end{equation}
where $g(x)$ is called the \textit{performance function}. The
performance of the system is measured by comparison of the
response $y$ with a specified critical value $y^*$: if $y\leq y^*$,
then the system is safe; if $y>y^*$, then the system has failed.
This failure criterion allows to define the \textit{failure domain}
$F$ in the input $x$-space as follows:
\begin{equation}\label{failure domain}
    F=\{x~:~g(x)>y^*\}.
\end{equation}
In other words, the failure domain is a set of values of input variables that lead to
unacceptance system performance, namely, to the exceedance of some
prescribed critical threshold $y^*$, which may represent the maximum
permissible roof displacement, maximum permissible interstory drift,
etc.

Engineering systems are complex systems, where complexity, in
particular, means that the information about the system (its
geometric and material properties) and its environment (loads) is
never complete. Therefore, there are always uncertainties in the
values of input variables $x$. To account for these uncertainties,
the input variables are modeled as random variables whose marginal
distributions are usually obtained from test data, expert opinion,
or from literature. Let $\pi(x)$ denote the join probability density
function (PDF) for $x$. The randomness in the input variables is
propagated through (\ref{performance function}) into the response
variable $y$, which makes the \textit{failure event} $\{x\in
F\}=\{y>y^*\}$ also random. The \textit{engineering reliability
problem} is then to compute the probability of failure $p_F$, given
by the following expression:
\begin{equation}\label{pF}
    p_F=\mathbb{P}(x\in F)=\int_F \pi(x)dx.
\end{equation}

The behavior of complex systems, such as tall buildings
and bridges, is represented by a complex model (\ref{performance
function}). In this context, complexity means that the performance
function $g(x)$, which defines the integration region $F$ in
(\ref{pF}), is not explicitly known. The evaluation of  $g(x)$ for
any $x$ is often time-consuming and usually done by the finite
element method (FEM), one of the most important numerical tools for
computation of the response of engineering systems. Thus, it is
usually impossible to evaluate the integral in (\ref{pF})
analytically because the integration region, the failure domain $F$,
is not known explicitly.

Moreover, traditional numerical integration is also generally not
applicable. In this approach, the $d$-dimensional input $x$-space is
partitioned into a union of disjoint hypercubes,
$\square_1,\ldots,\square_N$. For each hypercube $\square_i$, a
``representative'' point $x^{(i)}$ is chosen inside that hypercube,
$x^{(i)}\in\square_i$. The integral in (\ref{pF}) is then
approximated by the following sum:
\begin{equation}\label{numerical integration}
    p_F\approx \sum_{x^{(i)}\in
    F}\pi(x^{(i)})\mathrm{vol}(\square_i),
\end{equation}
where $\mathrm{vol}(\square_i)$ denotes the volume of $\square_i$
and summation is taken over all failure points $x^{(i)}$. Since it
is not known in advance whether a given point is a failure point or
not (the failure domain $F$ is not known explicitly), to compute the
sum in (\ref{numerical integration}), the failure criterion
(\ref{failure domain}) must be checked for all $x^{(i)}$. Therefore,
the approximation (\ref{numerical integration}) becomes
\begin{equation}\label{numerical integration2}
    p_F\approx\sum_{i=1}^NI_F(x^{(i)})\pi(x^{(i)})\mathrm{vol}(\square_i),
\end{equation}
where $I_F(x)$ stands for the indicator function, i.e.,
\begin{equation}\label{indicator}
    I_F(x)=\left\{
             \begin{array}{ll}
               1, & \hbox{if $x\in F$,} \\
               0, & \hbox{if $x\notin F$.}
             \end{array}
           \right.
\end{equation}

If $n$ denotes the number of intervals each dimension of the input
space is partitioned into, then the total number of terms in
(\ref{numerical integration2}) is $N=n^d$. Therefore, the
computational effort of numerical integration grows exponentially
with the number of dimensions $d$. In engineering reliability
problems, the dimension of the input space is typically very large
(e.g., when the stochastic load time history is discretized in
time). For example, $d\sim 10^3$ is not unusual in the reliability
literature. This makes numerical integration computationally
infeasible.

Over the past few decades, many different methods for solving the
engineering reliability problem (\ref{pF}) have been developed. In
general, the proposed reliability methods can be classified into
three categories, namely:
\begin{enumerate}
  \item[(a)] \textit{Analytic methods} are based on the Taylor-series
expansion of the performance function, e.g. the First-Order
Reliability Method (FORM) and the Second-Order Reliability Method
(SORM) \cite{Ditlevsen,Madsen,Melchers}.
  \item[(b)] \textit{Surrogate methods} are based on a functional surrogate of the performance
  function, e.g. the Response Surface Method (RSM)
  \cite{Faravelli,Schueller,Bucher}, Neural Networks \cite{Papadrakakis}, Support Vector
  Machines \cite{Hurtado}, and other methods \cite{Hurtado_book}.
  \item[(c)] \textit{Monte Carlo simulation methods}, among which
  are Importance Sampling \cite{Engelund}, Importance Sampling using Elementary Events \cite{AuBeck2},
  Radial-based Importance Sampling \cite{Grooteman}, Adaptive Linked Importance Sampling \cite{ALIS},
  Directional Simulation \cite{Ditlevsen}, Line Sampling \cite{LineSampling},
  Auxiliary Domain Method \cite{ADM}, Horseracing Simulation \cite{Zuev}, and
  \textit{Subset Simulation}~\cite{AuBeck}.
\end{enumerate}

Subset Simulation is thus a reliability method which is based on
(advanced) Monte Carlo simulation.

\section{The Direct Monte Carlo Method}\label{DMCsection}

The Monte Carlo method, referred in this paper as \textit{Direct
Monte Carlo} (DMC), is a statistical sampling technique that have
been originally developed by Stan Ulam, John von Neumann, Nick Metropolis (who
actually suggested the name ``Monte Carlo'' \cite{Metropolis}), and
their collaborators for solving the problem of neutron diffusion and
other problems in mathematical physics \cite{MonteCarlo}. From a
mathematical point of view, DMC allows to estimate the expected
value of a quantity of interest. More specifically, suppose the goal
is to evaluate $\mathbb{E}_\pi[h(x)]$, that is an expectation of a
function $h:\mathcal{X}\rightarrow\mathbb{R}$ with respect to the
PDF $\pi(x)$,
\begin{equation}\label{E}
    \mathbb{E}_{\pi}[h(x)]=\int_\mathcal{X} h(x)\pi(x)dx.
\end{equation}
The idea behind DMC is a straightforward application of the
\textit{law of large numbers} that states that if
$x^{(1)},x^{(2)},\ldots$ are i.i.d. (independent and identically
distributed) from the PDF $\pi(x)$, then the empirical average
$\frac{1}{N}\sum_{i=1}^Nh(x^{(i)})$ converges to the true value
$\mathbb{E}_{\pi}[h(x)]$ as $N$ goes to $+\infty$. Therefore, if the
number of samples $N$ is large enough, then $\mathbb{E}_{\pi}[h(x)]$
can be accurately estimated by the corresponding empirical average:
\begin{equation}\label{DMC}
     \mathbb{E}_{\pi}[h(x)]\approx \frac{1}{N}\sum_{i=1}^Nh(x^{(i)}).
\end{equation}

The relevance of DMC to the reliability problem (\ref{pF}) follows
from a simple observation that the failure probability $p_F$ can be
written as an expectation of the indicator function
(\ref{indicator}), namely,
\begin{equation}\label{pFasE}
     p_F=\int_F \pi(x)dx=\int_{\mathcal{X}}
I_F(x)\pi(x)dx=\mathbb{E}_{\pi}[I_F(x)],
\end{equation}
where $\mathcal{X}$ denotes the entire input $x$-space. Therefore,
the failure probability can be estimated using the DMC method
(\ref{DMC}) as follows:
\begin{equation}\label{pFDMC}
    p_F\approx\hat{p}_F^{\mbox{\tiny{DMC}}}=\frac{1}{N}\sum_{i=1}^NI_F(x^{(i)}),
\end{equation}
where $x^{(1)},\ldots x^{(N)}$ are i.i.d. samples from  $\pi(x)$.

The DMC estimate of $p_F$ is thus just the ratio of the total number
of \textit{failure samples} $\sum_{i=1}^NI_F(x^{(i)})$, i.e.,
samples that produce system failure according to the system model,
to the total number of samples, $N$. Note that
$\hat{p}_F^{\mbox{\tiny{DMC}}}$ is an \textit{unbiased} random
estimate of the failure probability, that is, on average,
$\hat{p}_F^{\mbox{\tiny{DMC}}}$  equals to $p_F$. Mathematically,
this means that $\mathbb{E}[\hat{p}_F^{\mbox{\tiny{DMC}}}]=p_F$.
Indeed, using the fact that $x^{(i)}\sim\pi(x)$ and (\ref{pFasE}),
\begin{equation}\label{DMCisUnbiased}
\begin{split}
\mathbb{E}[\hat{p}_F^{\mbox{\tiny{DMC}}}]&=\mathbb{E}\left[\frac{1}{N}\sum_{i=1}^NI_F(x^{(i)})\right]\\
&=\frac{1}{N}\sum_{i=1}^N\mathbb{E}[I_F(x^{(i)})]\\
&=\frac{1}{N}\sum_{i=1}^N\mathbb{E}_{\pi}[I_F(x)]=p_F.
\end{split}
\end{equation}

The main advantage of DMC over numerical integration is that its
accuracy does not depend on the dimension $d$ of the input space. In reliability analysis, the standard measure of accuracy of
an unbiased estimate $\hat{p}_F$ of the failure probability is its
\textit{coefficient of variation} (c.o.v.) $\delta(\hat{p}_F)$,
which is defined as the ratio of the standard deviation to the
expected value of $\hat{p}_F$, i.e.,
$\delta(\hat{p}_F)=\sqrt{\mathbb{V}[\hat{p}_F]}/\mathbb{E}[\hat{p}_F]$,
where $\mathbb{V}$ denotes the variance. The smaller the c.o.v.
$\delta(\hat{p}_F)$, the more accurate the estimate $\hat{p}_F$ is.
It is straightforward to calculate the variance of the DMC estimate:
\begin{equation}\label{V of DMC}
\begin{split}
\mathbb{V}[\hat{p}_F^{\mbox{\tiny{DMC}}}]&=\mathbb{V}\left[\frac{1}{N}\sum_{i=1}^NI_F(x^{(i)})\right]\\
&=
\frac{1}{N^2}\sum_{i=1}^N\mathbb{V}[I_F(x^{(i)})]\\
&=\frac{1}{N^2}\sum_{i=1}^N\left(\mathbb{E}[I_F(x^{(i)})^2]-\mathbb{E}[I_F(x^{(i)})]^2\right)\\
&=
\frac{1}{N^2}\sum_{i=1}^N\left(p_F-p_F^2\right)=\frac{p_F(1-p_F)}{N}.
\end{split}
\end{equation}
Here, the identity  $I_F(x)^2=I_F(x)$ was used. Using
(\ref{DMCisUnbiased}) and (\ref{V of DMC}), the c.o.v. of the DMC
estimate can be calculated:
\begin{equation}\label{cov_of_DMC}
  \delta(\hat{p}_F^{\mbox{\tiny{DMC}}})= \frac{\sqrt{\mathbb{V}[\hat{p}_F^{\mbox{\tiny{DMC}}}]}}{\mathbb{E}[\hat{p}_F^{\mbox{\tiny{DMC}}}]}
=\sqrt{\frac{1-p_F}{Np_F}}.
\end{equation}
This result shows that $\delta(\hat{p}_F^{\mbox{\tiny{DMC}}})$
depends only on the failure probability $p_F$ and the total number
of samples $N$, and does not depend on the dimension $d$ of the
input space. Therefore, unlike numerical integration, the DMC method
does not suffer from the ``curse of dimensionality'', i.e. from an exponential increase in volume associated
with adding extra dimensions, and is able to handle problems of high dimension.

Nevertheless, DMC has a serious drawback: it is inefficient in
estimating small failure probabilities. For typical engineering reliability problems, the failure probability $p_F$ is very small,
$p_F\ll1$. In other words, the system is usually assumed to be
designed properly, so that its failure is a \textit{rare event}. In
the reliability literature, $p_F\sim10^{-2}-10^{-9}$ have been
considered. If $p_F$ is very small, then it follows from
(\ref{cov_of_DMC}) that
\begin{equation}\label{cov of DMC 2}
    \delta(\hat{p}_F^{\mbox{\tiny{DMC}}})\approx\frac{1}{\sqrt{Np_F}}.
\end{equation}
This means that the number of samples $N$ needed to achieve an
acceptable level of accuracy is inverse proportional to $p_F$, and
therefore very large, $N\propto 1/p_F\gg1$. For example, if
$p_F=10^{-4}$ and the c.o.v. of $10\%$ is desirable, then $N=10^6$
samples are required. Note, however, that each evaluation of
$I_F(x^{(i)})$, $i=1,\ldots,N$, in (\ref{pFDMC}) requires a
system analysis to be performed to check whether the sample
$x^{(i)}$ is a failure sample. As it has been already mentioned in
Section \ref{Reliability Problem}, the computation effort for the
system analysis, i.e., computation of the performance function
$g(x)$, is significant (usually involves the FEM method). As a
result, the DMC method becomes excessively costly and practically
inapplicable for reliability analysis. This deficiency of DMC has
motivated research to develop more advanced simulation algorithms
for efficient estimation of small failure probabilities in
high dimensions.
\begin{remark} It is important to highlight, however, that even though DMC
cannot be routinely used for reliability problems (too
expensive), it is a very robust method, and it is often used as a
check on other reliability methods.
\end{remark}

\section{Preprocessing: Transformation of Input Variables}\label{Transformation}
Many reliability methods, including Subset Simulation, assume that
the input variables $x$ are independent. This assumption, however, is not a limitation,
since in simulation one always starts from independent variables to generate dependent ``physical'' variables.
Furthermore, for convenience, it is often assumed that $x$ are i.i.d. Gaussian.
If this is not the case, a ``preprocessing'' step that transforms $x$
to i.i.d. Gaussian variables $z$ must be undertaken. The transformation form $x$ to $z$ can be
performed in several ways depending on the available information
about the input variables. In the simplest case, when $x$ are
independent Gaussians, $x_k\sim\mathcal{N}(\cdot|\mu_k,\sigma_k^2)$,
where $\mu_k$ and $\sigma_k^2$ are respectively the mean and
variance of $x_k$, the necessary transformation is standardization:
\begin{equation}\label{standardization}
    z_k=\frac{x_k-\mu_k}{\sigma_k}.
\end{equation}
In other cases, more general techniques should be used, such as the
Rosenblatt transformation~\cite{Rosenblatt} and the Nataf
transformation~\cite{Nataf}. To avoid introduction of additional
notation, hereinafter, it is assumed without loss of generality that
the vector $x$ has been already transformed and it follows the
standard multivariate Gaussian distribution,
\begin{equation}\label{Gaussian}
    \pi(x_1,\ldots,x_d)=\prod_{k=1}^d \phi(x_k),
\end{equation}
where $\phi(\cdot)$ denotes the standard Gaussian PDF,
\begin{equation}\label{N(0,1)}
    \phi(x)=\frac{1}{\sqrt{2\pi}}\, e^{- \frac{\scriptscriptstyle 1}{\scriptscriptstyle 2} x^2}.
\end{equation}

\section{The Subset Simulation Method}\label{SS}

\begin{figure}[t]\centering
	\includegraphics[angle=0,scale=0.45]{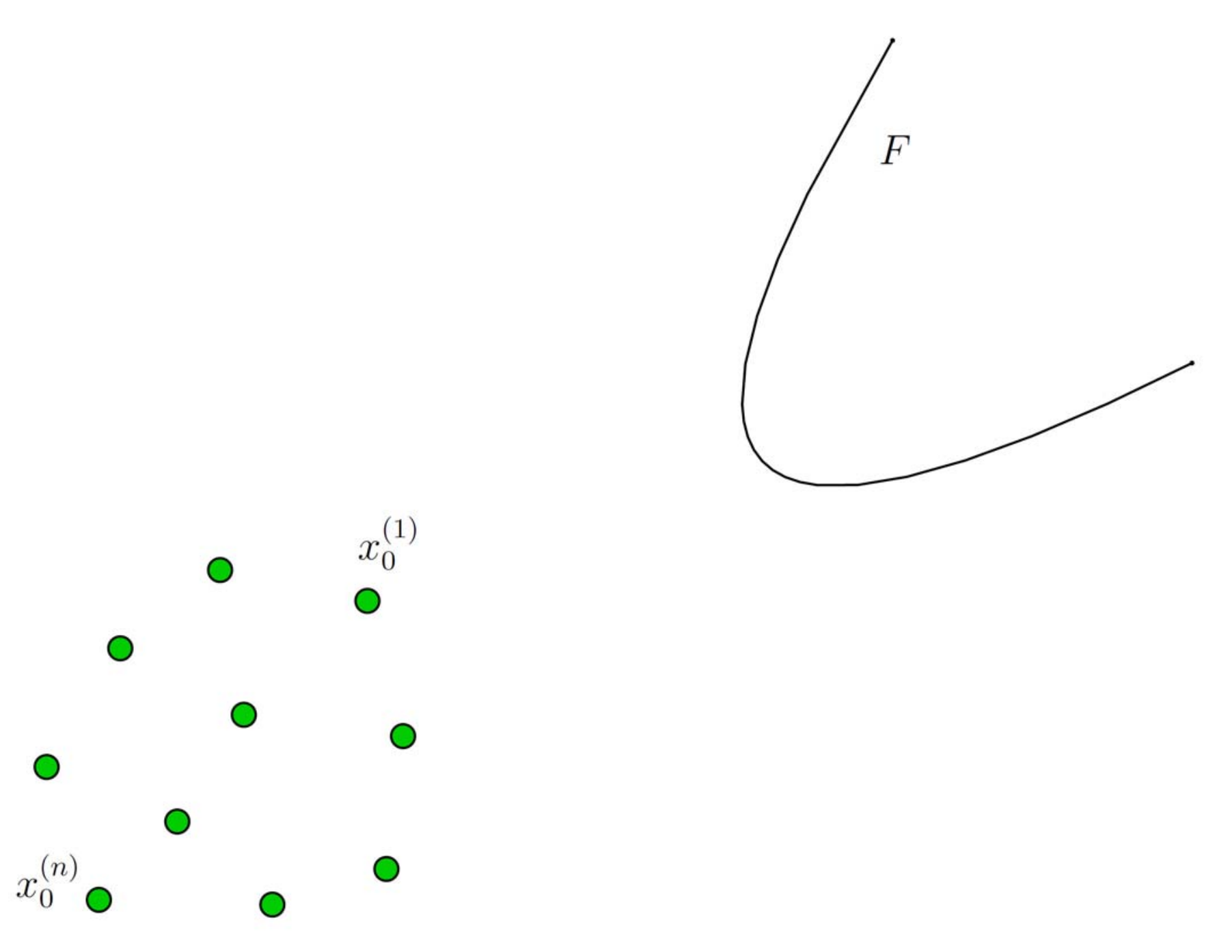}
	\caption{ Monte Carlo samples $x^{(1)}_0,\ldots,x^{(n)}_0$ and the failure domain $F$.
		$x^{(1)}_0$ and $x^{(n)}_0$ are, respectively, the closest to failure and the safest
		samples among $x^{(1)}_0,\ldots,x^{(n)}_0$. }  \label{fig1}
\end{figure}

Unlike Direct Monte Carlo, where all computational resources are
directly spent on sampling the input space,
$x^{(1)},\ldots,x^{(N)}\sim \pi(\cdot)$, and computing the values of
the performance function $g(x^{(1)}),\ldots,g(x^{(N)})$, Subset
Simulation first ``probes'' the input space $\mathcal{X}$ by generating a relatively
small number of i.i.d samples $x^{(1)}_0,\ldots,x^{(n)}_0\sim
\pi(x)$, $n<N$, and computing the corresponding system responses
$y_0^{(1)}=g(x_0^{(1)}),\ldots,y_0^{(n)}=g(x_0^{(n)})$. Here, the
subscript 0  indicates the $0^{\mathrm{th}}$ stage of the algorithm.
Since $F$ is a rare event and $n$ is relatively small, it is very
likely that none of the samples $x^{(1)}_0,\ldots,x^{(n)}_0$ belongs
to $F$, that is $y_0^{(i)}<y^*$ for all $i=1,\ldots,n$.
Nevertheless, these Monte Carlo samples contain some useful
information about the failure domain that can be utilized. To keep
the notation simple, assume that $y_0^{(1)},\ldots,y_0^{(n)}$ are
arranged in the decreasing order, i.e. $y_0^{(1)}\geq \ldots \geq
y_0^{(n)}$ (it is always possible to achieve this by renumbering
$x^{(1)}_0,\ldots,x^{(n)}_0$ if needed). Then, $x^{(1)}_0$ and
$x^{(n)}_0$ are, respectively, the closest to failure and the safest
samples among $x^{(1)}_0,\ldots,x^{(n)}_0$, since $y_0^{(1)}$ and
$y_0^{(n)}$ are the largest and the smallest responses. In general,
the smaller $i$, the closer to failure the sample $x_0^{(i)}$ is.
This is shown schematically in Fig.~\ref{fig1}.

\begin{figure}[t]\centering
	\includegraphics[angle=0,scale=0.45]{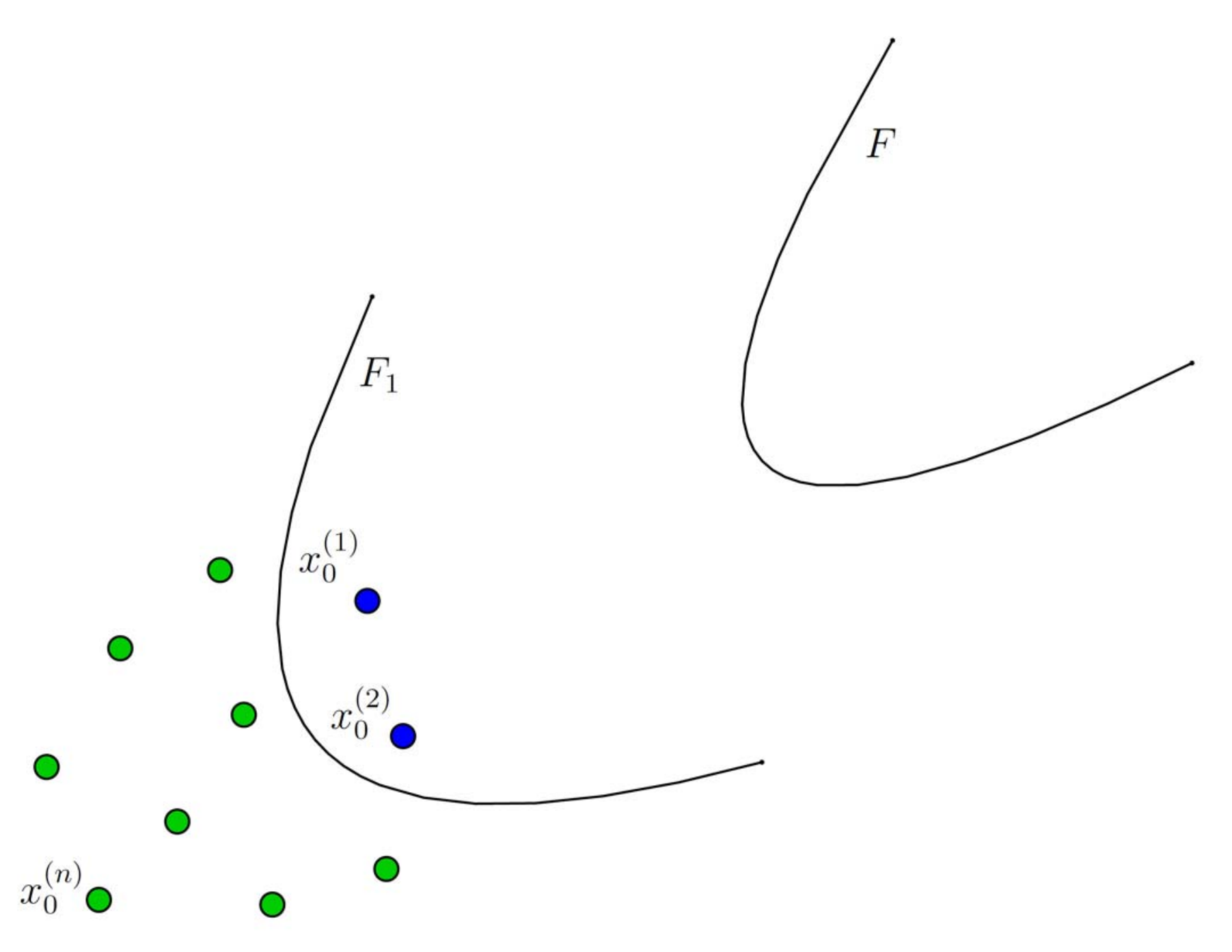}
	\caption{ The first intermediate failure domain $F_1$. In this schematic illustration, $n=10$, $p=0.2$, so that
		there are exactly $np=2$ Monte Carlo samples in $F_1$, $x_0^{(1)},x_0^{(2)}\in F_1$.}  \label{fig2}
\end{figure}

Let $p\in(0,1)$ be any number such that $np$ is integer. By analogy
with (\ref{failure domain}), define the \textit{first intermediate
failure domain} $F_1$ as follows:
\begin{equation}\label{F1dim2}
   F_1=\{x~:~g(x)>y_1^*\},
\end{equation}
where
\begin{equation}\label{y1*}
y_1^*=\frac{y_0^{(np)}+y_0^{(np+1)}}{2}.
\end{equation}
In other words, $F_1$ is the set of inputs that lead to the
exceedance of the \textit{relaxed threshold} $y^*_1<y^*$. Note that
by construction, samples $x^{(1)}_0,\ldots,x^{(np)}_0$ belong to
$F_1$, while $x^{(np+1)}_0,\ldots,x^{(n)}_0$ do not. As a
consequence, the Direct Monte Carlo estimate for the probability of
$F_1$ which is based on samples $x^{(1)}_0,\ldots,x^{(n)}_0$ is
automatically equal to $p$,
\begin{equation}\label{P(F1)}
    \mathbb{P}(F_1)\approx \frac{1}{n}\sum_{i=1}^n I_{F_1}(x^{(i)}_0)= p.
\end{equation}
The value $p=0.1$ is often used in the literature, which makes $F_1$
a relatively frequent event. Fig.~\ref{fig2} illustrates the
definition of $F_1$.

The first intermediate failure domain $F_1$ can be viewed as a (very
rough) conservative approximation to the target failure domain $F$.
Since $F\subset F_1$, the failure probability $p_F$ can be written
as a product:
\begin{equation}\label{pf=pf1p(f|f1)}
    p_F=\mathbb{P}(F_1)\mathbb{P}(F|F_1),
\end{equation}
where $\mathbb{P}(F|F_1)$ is the conditional probability of $F$
given  $F_1$. Therefore, in view of (\ref{P(F1)}), the problem of
estimating $p_F$ is reduced to estimating the conditional
probability $\mathbb{P}(F|F_1)$.

In the next stage, instead of generating samples in the whole input
space (like in DMC), the SS algorithm aims to populate $F_1$.
Specifically, the goal is to generate samples
$x^{(1)}_1,\ldots,x^{(n)}_1$ from the conditional distribution
\begin{equation}\label{pi(x|F1)}
    \pi(x|F_1)=\frac{\pi(x)I_{F_1}(x)}{\mathbb{P}(F_1)}=\frac{I_{F_1}(x)}{\mathbb{P}(F_1)}\prod_{k=1}^d \phi(x_k).
\end{equation}
First of all, note that samples $x^{(1)}_0,\ldots,x^{(np)}_0$ not
only belong to $F_1$, but are also distributed according to
$\pi(\cdot|F_1)$. To generate the remaining $(n-np)$ samples from
$\pi(\cdot|F_1)$, which, in general, is not a trivial task, Subset
Simulation uses the so-called \textit{Modified Metropolis algorithm}
(MMA). MMA belongs to the class of \textit{Markov chain Monte Carlo}
(MCMC ) algorithms~\cite{Liu,Robert}, which are techniques for
sampling from complex probability distributions that cannot be sampled
directly, at least not efficiently. MMA is based on the
original Metropolis algorithm \cite{MA} and specifically tailored
for sampling from the conditional distributions of the form
(\ref{pi(x|F1)}).

\subsection{Modified Metropolis algorithm}

Let $x\sim\pi(\cdot|F_1)$ be a sample from the conditional
distribution $\pi(\cdot|F_1)$. The Modified Metropolis algorithm
generates another sample $\tilde{x}$ from $\pi(\cdot|F_1)$ as
follows:
\begin{enumerate}
  \item Generate a ``candidate'' sample $\xi$: \\ For each coordinate $k=1,\ldots,d,$
  \begin{enumerate}
    \item Sample  $\eta_k\sim q_k(\cdot|x_k)$, where $q_k(\cdot|x_k)$, called the \textit{proposal distribution}, is a univariate
PDF for $\eta_k$ centered at $x_k$ with the symmetry property
$q_k(\eta_k|x_k)=q_k(x_k|\eta_k)$. For example, the proposal
distribution can be a Gaussian PDF with mean $x_k$ and variance
$\sigma^2_k$,
\begin{equation}\label{Gaussian proposal}
    q_k(\eta_k|x_k)=\frac{1}{\sqrt{2\pi}\sigma_k}\, \exp\left(- \frac{ (\eta_k-x_k)^2}{
    2\sigma_k^2}\right),
\end{equation}
or it can be a uniform distribution over $[x_k-\alpha,x_k+\alpha]$,
for some $\alpha\geq0$.
    \item Compute the acceptance ratio
\begin{equation}\label{AR}
    r_k=\frac{\phi(\eta_k)}{\phi(x_k)}.
\end{equation}
    \item Define the $k^{\mathrm{th}}$ coordinate of the candidate
    sample by accepting or rejecting $\eta_k$,
    \begin{equation}\label{xi_k}
    \xi_k=\left\{
                \begin{array}{ll}
                  \eta_k, & \hbox{with probability } \min\{1,r_k\}, \\
                  x_k, & \hbox{with probability } 1-\min\{1,r_k\}.
                \end{array}
              \right.
\end{equation}
  \end{enumerate}
  \item Accept or reject the candidate sample $\xi$ by setting
\begin{equation}\label{xtilda}
    \tilde{x}=\left\{
                \begin{array}{ll}
                  \xi, & \hbox{if } \xi\in F_1, \\
                  x, & \hbox{if } \xi\notin F_1.
                \end{array}
              \right.
\end{equation}
\end{enumerate}

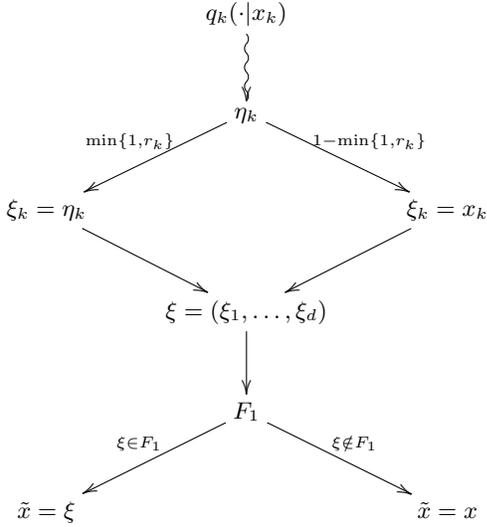
\begin{figure}[h]\centering
\[
\xymatrix{&& q_k(\cdot|x_k)\ar@{~>}[d] &&\\
&& \eta_k\ar[dl]_-{\min\{1,r_k\}}\ar[dr]^-{1-\min\{1,r_k\}} & &\\
&\xi_k=\eta_k\ar@{->}[dr]&& \xi_k=x_k\ar@{->}[dl]&& \\
&&\xi=(\xi_1,\ldots,\xi_d)\ar@{->}[d]&&&\\
&&F_1\ar[ld]_-{\xi\in F_1}\ar[rd]^-{\xi\notin F_1}&\\
&\tilde{x}=\xi&&\tilde{x}=x}
\]
\caption{Modified Metropolis algorithm} \label{M-alg}
\end{figure}

The Modified Metropolis algorithm is schematically illustrated in
Fig.~\ref{M-alg}.

It can be shown that the sample $\tilde{x}$ generated by MMA is
indeed distributed according to $\pi(\cdot|F_1)$. If the candidate
sample $\xi$ is rejected in (\ref{xtilda}), then
$\tilde{x}=x\sim\pi(\cdot|F_1)$ and there is nothing to prove.
Suppose now that $\xi$ is accepted, $\tilde{x}=\xi$, so that the
move from $x$ to $\tilde{x}$ is a proper transition between two
distinct points in $F_1$. Let $f(\cdot)$ denote the PDF of
$\tilde{x}$ (the goal is to show that
$f(\tilde{x})=\pi(\tilde{x}|F_1)$). Then
\begin{equation}\label{f(xtilda)}
    f(\tilde{x})=\int_{F_1} \pi(x|F_1)t(\tilde{x}|x)dx,
\end{equation}
where $t(\tilde{x}|x)$ is the transition PDF from $x$ to
$\tilde{x}\neq x$. According to the first step of MMA, coordinates
of $\tilde{x}=\xi$ are generated independently, and therefore
$t(\tilde{x}|x)$ can be expressed as a product,
\begin{equation}\label{t_product}
t(\tilde{x}|x)=\prod_{k=1}^dt_k(\tilde{x}_k|x_k),
\end{equation}
where $t_k(\tilde{x}_k|x_k)$ is the transition PDF for the
$k^{\mathrm{th}}$ coordinate $\tilde{x}_k$. Combining equations
(\ref{pi(x|F1)}), (\ref{f(xtilda)}), and (\ref{t_product}) gives
\begin{equation}\label{f(xtilda2)}
\begin{split}
f(\tilde{x})&=\int_{F_1} \frac{I_{F_1}(x)}{\mathbb{P}(F_1)}\prod_{k=1}^d \phi(x_k)\prod_{k=1}^dt_k(\tilde{x}_k|x_k)dx\\
&=
\frac{1}{\mathbb{P}(F_1)}\int_{F_1}\prod_{k=1}^d\phi(x_k)t_k(\tilde{x}_k|x_k)dx.
\end{split}
\end{equation}
The key to the proof of $f(\tilde{x})=\pi(\tilde{x}|F_1)$ is to
demonstrate that $\phi(x_k)$ and $t_k(\tilde{x}_k|x_k)$ satisfy the
so-called \textit{detailed balance equation},
\begin{equation}\label{dbe}
    \phi(x_k)t_k(\tilde{x}_k|x_k)=\phi(\tilde{x}_k)t_k(x_k|\tilde{x}_k).
\end{equation}
If $\tilde{x}_k=x_k$, then (\ref{dbe}) is trivial. Suppose that
$\tilde{x}_k\neq x_k$, that is $\tilde{x}_k=\xi_k=\eta_k$ in
(\ref{xi_k}). The actual transition PDF $t_k(\tilde{x}_k|x_k)$ from
$x_k$ to $\tilde{x}_k\neq x_k$ differs from the proposal PDF
$q_k(\tilde{x}_k|x_k)$ because the acceptance-rejection step
(\ref{xi_k}) is involved. To actually make the move from $x_k$ to
$\tilde{x}_k$, one needs not only to generate $\tilde{x}_k\sim
q_k(\cdot|x_k)$, but also to accept it with probability
$\min\{1,\frac{\phi(\tilde{x}_k)}{\phi(x_k)}\}$. Therefore,
\begin{equation}\label{transitionPDFk}
    t_k(\tilde{x}_k|x_k)=q_k(\tilde{x}_k|x_k)\min\left\{1,\frac{\phi(\tilde{x}_k)}{\phi(x_k)}\right\},
\hspace{3mm}  \tilde{x}_k\neq x_k.
\end{equation}
Using (\ref{transitionPDFk}), the symmetry property of the proposal
PDF, $q_k(\tilde{x}_k|x_k)=q_k(x_k|\tilde{x}_k)$, and the identity
$a\min\{1,\frac{b}{a}\}=b\min\{1,\frac{a}{b}\}$ for any $a,b>0$,
\begin{equation}\label{dbe2}
\begin{split}
  \phi(x_k)t_k(\tilde{x}_k|x_k)= & q_k(\tilde{x}_k|x_k)\phi(x_k)\min\left\{1,\frac{\phi(\tilde{x}_k)}{\phi(x_k)}\right\} \\
   =
   &q_k(x_k|\tilde{x}_k)\phi(\tilde{x}_k)\min\left\{1,\frac{\phi(x_k)}{\phi(\tilde{x}_k)}\right\}\\
   =&\phi(\tilde{x}_k)t_k(x_k|\tilde{x}_k),
\end{split}
\end{equation}
and the detailed balance (\ref{dbe}) is thus established. The rest
is a straightforward calculation:
\begin{equation}\label{f=final}
\begin{split}
f(\tilde{x})&=  \frac{1}{\mathbb{P}(F_1)}\int_{F_1}\prod_{k=1}^d\phi(\tilde{x}_k)t_k(x_k|\tilde{x}_k)dx\\
&=
\frac{1}{\mathbb{P}(F_1)}\prod_{k=1}^d\phi(\tilde{x}_k)\int_{F_1}t(x|\tilde{x})dx=\pi(\tilde{x}|F_1),
\end{split}
\end{equation}
 since the transition PDF $t(x|\tilde{x})$ integrates to $1$, and
$I_{F_1}(\tilde{x})=1$.

\begin{remark}
A mathematically more rigorous proof of the Modified Metropolis
algorithm is given in \cite{BSS}.
\end{remark}

\begin{remark}
It is worth mentioning that although the independence of input
variables is crucial for the applicability of MMA, and thus for
Subset Simulation, they need not be identically distributed. In
other words, instead of $(\ref{Gaussian})$, the joint PDF
$\pi(\cdot)$ can have a more general form, $\pi(x)=\prod_{k=1}^d
\pi_k(x_k)$, where $\pi_k(\cdot)$ is the marginal distributions of
$x_k$ which is not necessarily Gaussian. In this case, the
expression for the acceptance ratio in (\ref{AR}) must be replaced
by $r_k=\frac{\pi_k(\eta_k)}{\pi_k(x_k)}$.
\end{remark}

\subsection{Subset Simulation at higher conditional levels}

Given $x^{(1)}_0,\ldots,x^{(np)}_0\sim\pi(\cdot|F_1)$, it is clear
now how to generate the remaining $(n-np)$ samples from
$\pi(\cdot|F_1)$. Namely, starting from each $x^{(i)}_0$,
$i=1,\ldots, np$, the SS algorithm generates a sequence of
$(1-\frac{1}{p})$ new MCMC  samples $x^{(i)}_0=x^{(i)}_{0,0} \mapsto
x^{(i)}_{0,1} \mapsto \ldots \mapsto x^{(i)}_{0,1-\frac{1}{p}}$
using the Modified Metropolis transition rule described above. Note
that when $x^{(i)}_{0,j}$ is generated, the previous sample
$x^{(i)}_{0,j-1}$ is used as an input for the transition rule. The
sequence $x^{(i)}_{0,0}, x^{(i)}_{0,1}, \ldots,
x^{(i)}_{0,1-\frac{1}{p}}$ is called a \textit{Markov chain} with
the \textit{stationary distribution} $\pi(\cdot|F_1)$, and
$x^{(i)}_{0,0}=x^{(i)}_0$ is often referred to as the ``seed'' of
the Markov chain.

\begin{figure}[t]\centering
\includegraphics[angle=0,scale=0.45]{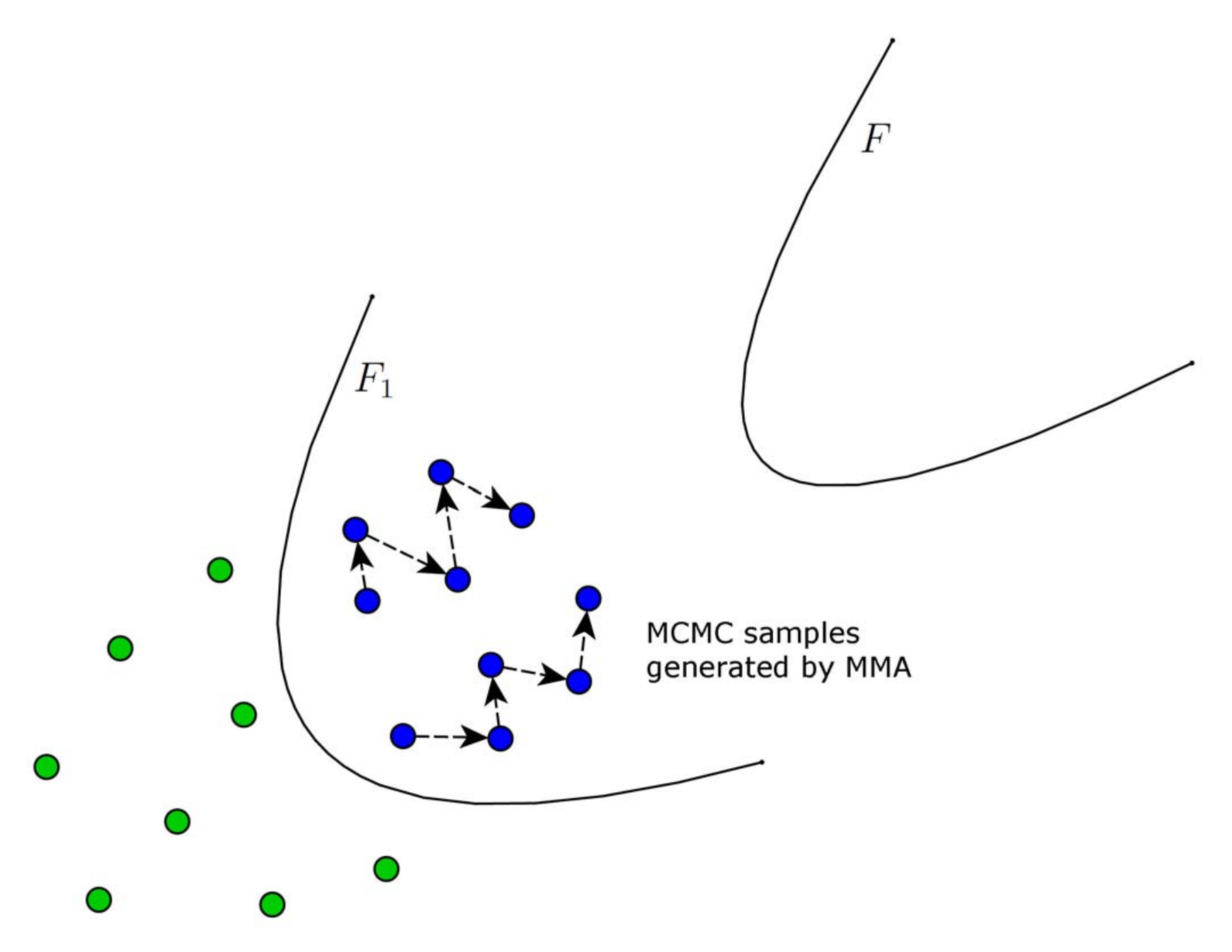}
\caption{ MCMC samples generated by the Modified Metropolis algorithm at the first conditional level of Subset Simulation.}  \label{fig4}
\end{figure}

To simplify the notation, denote samples
$\{x_{0,j}^{(i)}\}_{j=0,\ldots,1-\frac{1}{p}}^{i=1,\ldots,np}$
 by
$\{x^{(1)}_1,\ldots,x^{(n)}_1\}$. The subscript 1 indicates that the
MCMC samples $x^{(1)}_1,\ldots,x^{(n)}_1\sim\pi(\cdot|F_1)$ are
generated at the first conditional level of the SS algorithm. These
conditional samples are schematically shown in Fig.~\ref{fig4}.
Also assume that the corresponding system responses
$y_1^{(1)}=g(x^{(1)}_1),\ldots, y^{(n)}_1=g(x^{(n)}_1)$ are arranged
in the decreasing order, i.e. $y_1^{(1)}\geq\ldots\geq y^{(n)}_1$.
If the failure event $F$ is rare enough, that is if $p_F$ is
sufficiently small, then it is very likely that none of the samples
$x^{(1)}_1,\ldots,x^{(n)}_1$ belongs to $F$, i.e. $y_1^{(i)}<y^*$
for all $i=1,\ldots,n$. Nevertheless, these MCMC samples can be used
in the similar way the Monte Carlo samples
$x^{(1)}_0,\ldots,x^{(n)}_0$ were used.

\begin{figure}[h]\centering
\includegraphics[angle=0,scale=0.45]{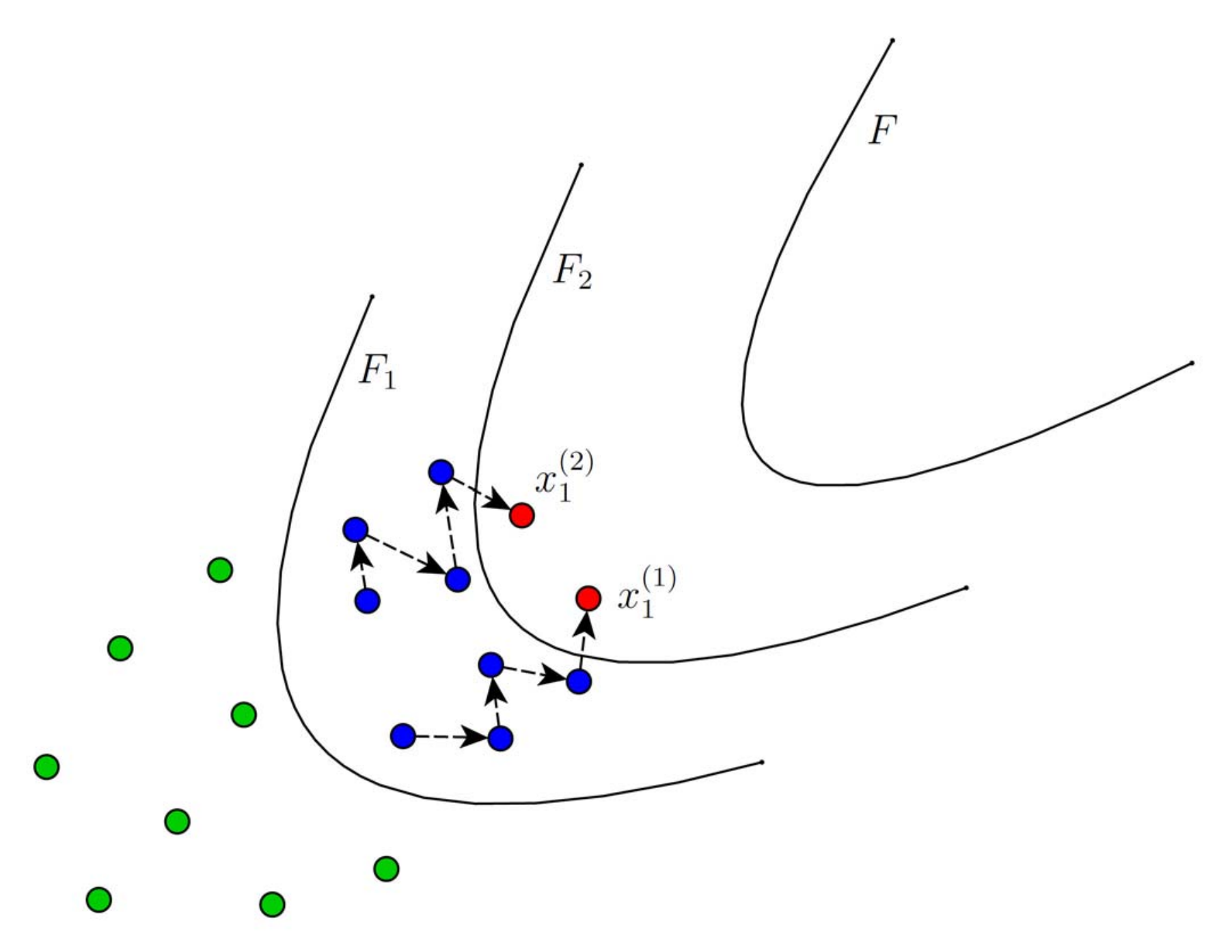}
\caption{\footnotesize The second intermediate failure domain $F_2$. In this schematic illustration, $n=10$, $p=0.2$, so that
there are exactly $np=2$ MCMC samples in $F_2$, $x_1^{(1)},x_1^{(2)}\in F_2$.}  \label{fig5}
\end{figure}

By analogy with (\ref{F1dim2}), define the \textit{second
intermediate failure domain} $F_2$ as follows:
\begin{equation}\label{F2dim2}
   F_2=\{x~:~g(x)>y_2^*\}, 
\end{equation}
where
\begin{equation}\label{y2*}
y_2^*=\frac{y_1^{(np)}+y_1^{(np+1)}}{2}.
\end{equation}
Note that $y_2^*>y_1^*$ since $y_1^{(i)}>y_1^*$ for all
$i=1,\ldots,n$. This means that $F\subset F_2\subset F_1$, and
therefore, $F_2$ can be viewed as a conservative approximation to
$F$ which is still rough, yet more accurate than $F_1$.
Fig.~\ref{fig5} illustrates the definition of $F_2$.
By construction,
samples $x^{(1)}_1,\ldots,x^{(np)}_1$ belong to $F_2$, while
$x^{(np+1)}_1,\ldots,x^{(n)}_1$ do not. As a result, the estimate
for the conditional probability of $F_2$ given $F_1$ which is based
on samples $x^{(1)}_1,\ldots,x^{(n)}_1\sim\pi(\cdot|F_1)$ is
automatically equal to $p$,
\begin{equation}\label{P(F2|F1)}
    \mathbb{P}(F_2|F_1)\approx \frac{1}{n}\sum_{i=1}^n I_{F_2}(x^{(i)}_1)=p.
\end{equation}

Since $F\subset F_2\subset F_1$, the conditional probability
$\mathbb{P}(F|F_1)$ that appears in (\ref{pf=pf1p(f|f1)}) can be
expressed as a product:
\begin{equation}\label{p(f|f1)=}
    \mathbb{P}(F|F_1)=\mathbb{P}(F_2|F_1)\mathbb{P}(F|F_2).
\end{equation}
Combining (\ref{pf=pf1p(f|f1)}) and (\ref{p(f|f1)=}) gives the
following expression for the failure probability:
\begin{equation}\label{pf=pf1pf2pf}
    p_F=\mathbb{P}(F_1)\mathbb{P}(F_2|F_1)\mathbb{P}(F|F_2).
\end{equation}
Thus, in view of (\ref{P(F1)}) and (\ref{P(F2|F1)}), the problem of
estimating $p_F$ is now reduced to estimating the conditional
probability $\mathbb{P}(F|F_2)$.

In the next step, as one may have already guessed, the Subset
Simulation algorithm: populates $F_2$ by generating MCMC samples
$x^{(1)}_2,\ldots,x^{(n)}_2$ from $\pi(\cdot|F_2)$ using the
Modified Metropolis algorithm; defines the third intermediate
failure domain $F_3\subset F_2$ such that
$\mathbb{P}(F_3|F_2)\approx\frac{1}{n}\sum_{i=1}^n
I_{F_3}(x^{(i)}_2)=p$ ; and reduces the original problem of
estimating the failure probability $p_F$ to estimating the
conditional probability $\mathbb{P}(F|F_3)$ by representing
$p_F=\mathbb{P}(F_1)\mathbb{P}(F_2|F_1)\mathbb{P}(F_3|F_2)\mathbb{P}(F|F_3)$.
The algorithm proceeds in this way until the target failure domain
$F$ has been sufficiently sampled so that the conditional
probability $\mathbb{P}(F|F_L)$ can be accurately estimated by
$\frac{1}{n}\sum_{i=1}^n I_{F}(x^{(i)}_L)$, where $F_L$ is the
$L^{\mathrm{th}}$ intermediate failure domain, and
$x^{(1)}_L,\ldots,x^{(n)}_L\sim\pi(\cdot|F_L)$ are the MCMC samples
generated at the $L^{\mathrm{th}}$ conditional level. Subset
Simulation can thus be viewed as a method that decomposes the rare
failure event $F$ into a sequence of progressively ``less-rare''
nested events, $F\subset F_L\subset\ldots\subset F_1$, where all
intermediate failure events $F_1,\ldots,F_L$ are constructed
adaptively by appropriately relaxing the value of the critical
threshold $y_1^*<\ldots<y_L^*<y^*$.

\subsection{Stopping criterion}

In what follows, the stopping criterion for Subset Simulation is
described in detail. Let $n_F(l)$ denote the number of failure
samples at the $l^{\mathrm{th}}$ level, that is
\begin{equation}\label{nj}
    n_F(l)=\sum_{i=1}^n I_F(x_l^{(i)}),
\end{equation}
where $x_l^{(1)},\ldots,x_l^{(n)}\sim\pi(\cdot|F_l)$. Since $F$ is a
rare event, it is very likely that $n_F(l)=0$ for the first few
conditional levels. As $l$ gets larger, however, $n_F(l)$ starts
increasing since $F_l$, which approximates $F$ ``from above'',
shrinks closer to $F$. In general, $n_{F}(l)\geq n_{F}(l-1)$, since
$F\subset F_{l}\subset F_{l-1}$ and the $np$ closest to $F$ samples
among $x_{l-1}^{(1)},\ldots,x_{l-1}^{(n)}$ are present among
$x_l^{(1)},\ldots,x_l^{(n)}$. At conditional level $l$, the failure
probability $p_F$ is expressed as a product,
\begin{equation}\label{pFLevelj}
    p_F=\mathbb{P}(F_1)\mathbb{P}(F_{2}|F_{1})\ldots\mathbb{P}(F_l|F_{l-1})\mathbb{P}(F|F_l).
\end{equation}
Furthermore, the adaptive choice of intermediate critical thresholds
$y_1^*,\ldots,y_l^*$ guarantees that the first $l$ factors in
(\ref{pFLevelj}) approximately equal to $p$, and, thus,
\begin{equation}\label{pFLeveljapprox}
    p_F\approx p^l\cdot\mathbb{P}(F|F_l).
\end{equation}
Since there are exactly $n_F(l)$ failure samples at the
$l^{\mathrm{th}}$ level, the estimate of the last conditional
probability in (\ref{pFLevelj}) which is based on samples
$x_l^{(1)},\ldots,x_l^{(n)}\sim\pi(\cdot|F_l)$ is given by
\begin{equation}\label{lastestimate}
\mathbb{P}(F|F_l)\approx \frac{1}{n}\sum_{i=1}^n
I_F(x_l^{(i)})=\frac{n_F(l)}{n}.
\end{equation}
If $n_F(l)$ is sufficiently large, i.e. the conditional event
$(F|F_l)$ is not rare, then the estimate (\ref{lastestimate}) is
fairly accurate. This leads to the following stopping criterion:
\begin{itemize}
  \item If $\frac{n_F(l)}{n}\geq p$, i.e. there are at least $np$
failure samples among $x_l^{(1)},\ldots,x_l^{(n)}$, then Subset
Simulation stops: the current conditional level $l$ becomes the last
level, $L=l$, and the failure probability estimate derived from
(\ref{pFLeveljapprox}) and (\ref{lastestimate}) is
\begin{equation}\label{pf estimate final}
    p_F\approx \hat{p}_F^{\mbox{\tiny{SS}}}=p^L\frac{n_F(L)}{n}.
\end{equation}
  \item If $\frac{n_F(l)}{n}<p$, i.e. there are less than $np$
failure samples among $x_l^{(1)},\ldots,x_l^{(n)}$, then the
algorithm proceeds by defining the next intermediate failure domain
$F_{l+1}=\{x:g(x)>y_{l+1}^*\}$, where
$y_{l+1}^*=(y_l^{(np)}+y_l^{(np+1)})/2$, and expressing
$\mathbb{P}(F|F_l)$ as a product
$\mathbb{P}(F|F_l)=\mathbb{P}(F_{l+1}|F_l)\mathbb{P}(F|F_{l+1})\approx
p\cdot\mathbb{P}(F|F_{l+1})$.
\end{itemize}

The described stopping criterion guarantees that the estimated
values of all factors in the factorization
$p_F=\mathbb{P}(F_1)\mathbb{P}(F_{2}|F_{1})\ldots\mathbb{P}(F_L|F_{L-1})\mathbb{P}(F|F_L)$
are not smaller than $p$. If $p$ is relatively large ($p=0.1$ is
often used in applications), then it is likely that the estimates
$\mathbb{P}(F_1)\approx p$, $\mathbb{P}(F_{2}|F_{1})\approx p, \ldots, \mathbb{P}(F_L|F_{L-1})\approx p$, and
$\mathbb{P}(F|F_L)\approx \frac{n_F(L)}{n}(\geq p)$ are accurate
even when the sample size $n$ is relatively small. As a result, the
SS estimate (\ref{pf estimate final}) is also accurate in this case.
This provides an intuitive explanation as to why Subset Simulation
is efficient in estimating small probabilities of rare events.
For a detailed discussion of error estimation for the SS method the reader is referred to \cite{SSbook}.

\subsection{Implementation details}

In the rest of this section, the implementation details of Subset
Simulation are discussed. The SS algorithm has two essential
components that affect its efficiency: the parameter $p$ and the set
of univariate proposal PDFs $\{q_k\}$, $k=1,\ldots,d$.

\subsubsection{Level probability}
The parameter
$p$, called  the \textit{level probability} in \cite{SSbook} and the
\textit{conditional failure probability} in \cite{BSS}, governs how
many intermediate failure domains $F_l$ are needed to reach the
target failure domain $F$. As it follows form (\ref{pf estimate
final}), a small value of $p$ leads to a fewer total number of
conditional levels $L$. But at the same time, it results in a large
number of samples $n$ needed at each conditional level $l$ for
accurate determination of $F_l$ (i.e. determination of $y_l^*$) that
satisfies $\frac{1}{n}\sum_{i=1}^nI_{F_l}(x^{(i)}_{l-1})=p$. In the
extreme case when $p\leq p_F$, no levels are needed, $L=0$, and
Subset Simulation reduces to the Direct Monte Carlo method. On the
other hand, increasing the value of $p$ will mean that fewer samples
are needed at each conditional level, but it will increase the total
number of  levels $L$. The choice of the level probability $p$ is
thus a tradeoff between the total number of level $L$ and the number
of samples $n$ at each level. In the original paper \cite{AuBeck},
it has been found that the value $p=0.1$ yields good efficiency. The
latter studies \cite{SSbook,BSS}, where the c.o.v. of the SS
estimate $\hat{p}_F^{\mbox{\tiny{SS}}}$ has been analyzed, confirmed that $p=0.1$
is a nearly optimal value of the level probability.

\subsubsection{Proposal distributions}

The efficiency and accuracy of Subset Simulation also depends on the
set of univariate proposal PDFs $\{q_k\}$, $k=1,\ldots,d$ that are
used within the Modified Metropolis algorithm for sampling from the
conditional distributions $\pi(\cdot|F_l)$. To see this, note that
in contract to the Monte Carlo samples
$x_0^{(1)},\ldots,x_0^{(n)}\sim\pi(\cdot)$ which are i.i.d., the
MCMC samples $x_l^{(1)},\ldots,x_l^{(n)}\sim\pi(\cdot|F_l)$ are
\textit{not independent} for $l\geq1$, since the MMA transition rule
uses $x_l^{(i)}\sim\pi(\cdot|F_l)$ to generate
$x_l^{(i+1)}\sim\pi(\cdot|F_l)$. This means that although  these
MCMC samples can be used for statistical averaging as if they were
i.i.d., the efficiency of the averaging is reduced if compared with
the i.i.d. case \cite{Doob}. Namely, the more correlated
$x_l^{(1)},\ldots,x_l^{(n)}$ are, the slower is the convergence of
the estimate
$P(F_{l+1}|F_l)\approx\frac{1}{n}\sum_{i=1}^nI_{F_{l+1}}(x_l^{(i)})$,
and, therefore, the less efficient it is. The correlation between
samples $x_l^{(1)},\ldots,x_l^{(n)}$ is due to proposal PDFs
$\{q_k\}$, which govern the generation of the next sample
$x_l^{(i+1)}$ from the current one $x_l^{(i)}$. Hence, the choice of
$\{q_k\}$ is very important.

It was observed in \cite{AuBeck} that the efficiency of MMA is not
sensitive to the type of the proposal PDFs (Gaussian, uniform, etc),
however, it strongly depends on their \textit{spread} (variance).
Both small and large spreads tend to increase the correlation
between successive samples. Large spreads may reduce the acceptance
rate in (\ref{xtilda}), increasing the number of repeated MCMC
samples. Small spreads, on the contrary, may lead to a reasonably
high acceptance rate, but still produce very correlated samples due
to their close proximity.  As a rule of thumb, the spread of $q_k$,
$k=1,\dots,d$, can be taken of the same order as the spread of the
corresponding marginal PDF $\pi_k$ \cite{SSbook}. For example, if
$\pi$ is given by (\ref{Gaussian}), so that all marginal PDFs are
standard Gaussian, $\pi_k(x)=\phi(x)$, then all proposal PDFs can
also be Gaussian with unit variance, $q_k(x|x_k)=\phi(x-x_k)$. This
choice is found to give a balance between efficiency and robustness.

The spread of proposal PDFs can also be chosen adaptively. In
\cite{BSS}, where the problem of optimal scaling for the Modified
Metropolis algorithm was studied in more detail, the following
nearly optimal scaling strategy was proposed: at each conditional
level, select the spread such that the the corresponding acceptance
rate in (\ref{xtilda}) is between $30\%$ and $50\%$. In general,
finding the optimal spread of proposal distributions is problem
specific and a highly non-trivial task not only for MMA, but also
for almost all MCMC algorithms.

\section{Illustrative Examples}\label{Examples}

To illustrate Subset Simulation and to demonstrate its efficiency in
estimating small probabilities of rare failure events, two examples
are considered in this section. As it has been discussed in Section
\ref{Reliability Problem}, in reliability problems, the dimension
$d$ of the input space $\mathcal{X}$ is usually very large. In spite
of this, for visualization and educational purposes, a linear
reliability problem in two dimensions ($d=2$) is first considered in
Section \ref{ex1}. A more realistic high-dimensional example
($d=10^3$) is considered in the subsequent Section \ref{ex2}.

\subsection{Subset Simulation in 2-D} \label{ex1}

Suppose that $d=2$, i.e. the
response variable $y$ depends only on two input variables $x_1$ and
$x_2$. Consider a linear performance function,
\begin{equation}\label{linear perf func 2d}
    g(x_1,x_2)=x_1+x_2,
\end{equation}
where  $x_1$ and $x_2$ are independent standard Gaussian,
$x_i\sim\mathcal{N}(0,1)$, $i=1,2.$ The failure domain $F$ is then a
half-plane defined by
\begin{equation}\label{Flinear2d}
    F=\{(x_1,x_2)~:~x_1+x_2>y^*\}.
\end{equation}

In this example, the failure probability $p_F$ can be calculated
analytically. Indeed, since $x_1+x_2\sim \mathcal{N}(0,2)$, and,
therefore, $\frac{x_1+x_2}{\sqrt{2}}\sim \mathcal{N}(0,1)$,
\begin{equation}\label{pftrue2d}
\begin{split}
p_F&=\mathbb{P}(x_1+x_2>y^*)=\mathbb{P}\left(\frac{x_1+x_2}{\sqrt{2}}>\frac{y^*}{\sqrt{2}}\right)\\
&=1-\Phi\left(\frac{y^*}{\sqrt{2}}\right),
\end{split}
\end{equation}
where $\Phi$ is the standard Gaussian CDF. This expression for the
failure probability can be used as a check on the SS estimate.
Moreover, expressing $y^*$ in terms of $p_F$,
\begin{equation}\label{y*2d}
    y^*=\sqrt{2}\Phi^{-1}(1-p_F),
\end{equation}
allows to solve the inverse problem, namely, to formulate a linear reliability problem with a given value of the failure probability.
Suppose that $p_F=10^{-10}$ is the target value. Then the corresponding value of the critical threshold
is $y^*\approx9$.

Subset Simulation were used to estimate the failure probability of the rare event (\ref{Flinear2d}) with $y^*=9$.
The parameters of the algorithm were chosen as follows:
the level probability $p=0.1$, the proposal PDFs $q_k(x|x_k)=\phi(x-x_k)$, and the sample size $n=10^3$ per each level.
This implementation of SS led to $L=9$ conditional levels,
making the total number of generated samples $N=n+L(n-np)=9.1\times10^3$.
The obtained SS estimate is $\hat{p}_F^{\mbox{\tiny{SS}}}=1.58\times10^{-10}$ which is quite close to the true value $p_F=10^{-10}$.
Note that, in this example, it is hopeless to obtain an accurate estimate by the Direct Monte Carlo method since
the DMC estimate (\ref{pFDMC}) based on $N=9.1\times10^3$ samples is effectively zero: the rare event $F$ is too rare.

\begin{figure}[t]\centering
\includegraphics[angle=0,scale=0.47]{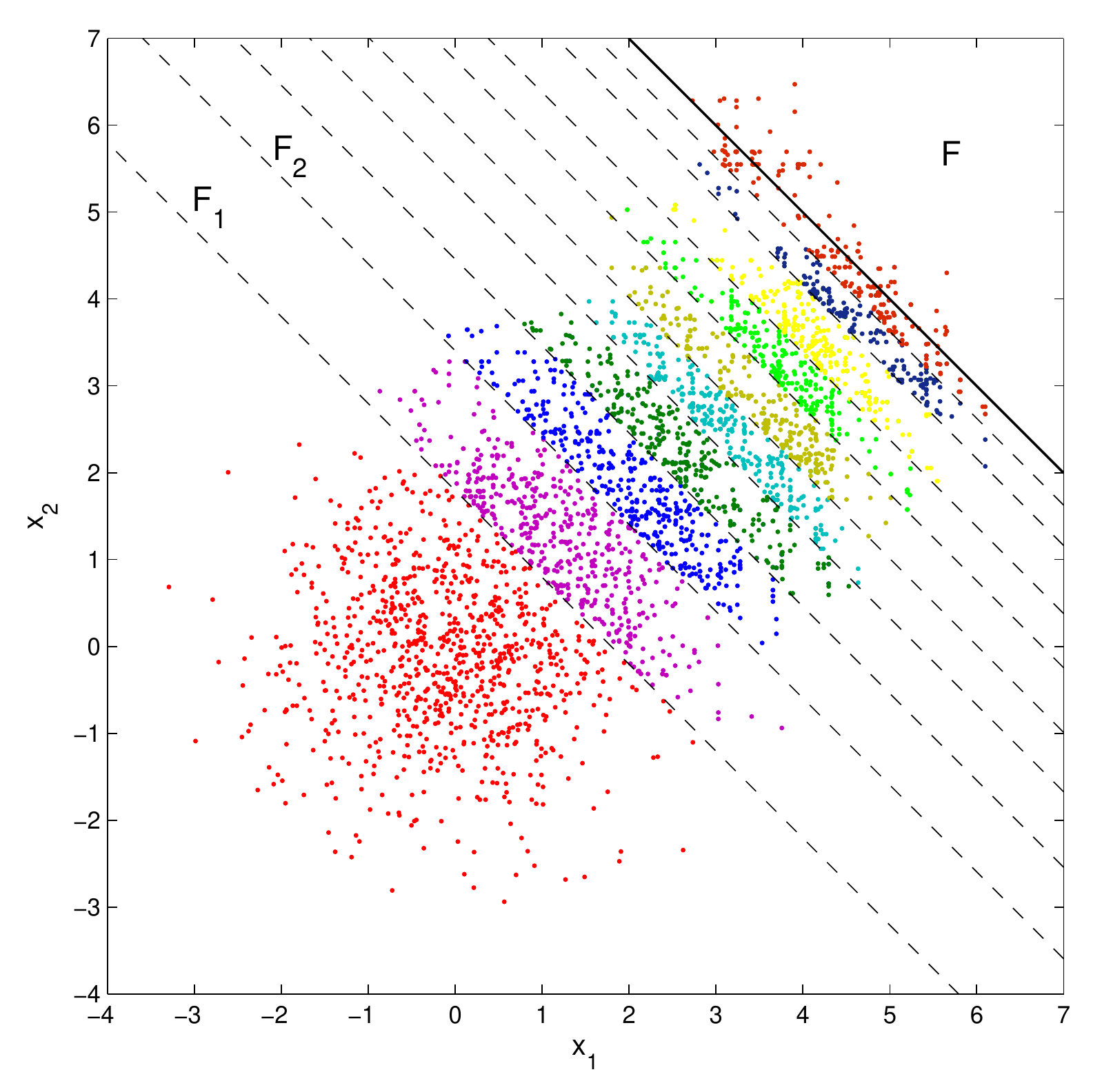}
\caption{Samples generated by Subset Simulation: red samples are Monte Carlo samples generated at the $0^{\mathrm{th}}$ unconditional level,
purple sample are MCMC sample generated at the $1^{\mathrm{st}}$ conditional level, etc. The dashed lines represent the boundaries of intermediate failure domains $F_l$, $l=1,\ldots,L=9$.
The solid line is the boundary of the target failure domain $F$.
[Example 6.1].} \label{FigA0}
\end{figure}

Fig.~\ref{FigA0} shows the samples generated by the SS method.
The dashed lines represent the boundaries of intermediate failure domains $F_l$, $l=1,\ldots,L=9$.
The solid line is the boundary of the target failure domain $F$. This illustrates how Subset Simulation pushes Monte Carlo samples (red)
towards the failure region.

\subsection{Subset Simulation in High Dimensions}\label{ex2}

It is straightforward to generalize the low-dimensional example considered in the previous section to high dimensions.
Consider a linear performance function
\begin{equation}\label{perf_func_d}
    g(x)=\sum_{i=1}^dx_i,
\end{equation}
where $x_1,\ldots,x_d$ are i.i.d. standard Gaussian. The failure domain is then a half-space defined by
\begin{equation}\label{Flineard}
      F=\{x~:~\sum_{i=1}^dx_i>y^*\}.
\end{equation}
In this example, $d=10^3$ is considered, hence the input space $\mathcal{X}=\mathbb{R}^d$ is indeed high-dimensional.
As before, the failure probability can be calculated analytically:
\begin{equation}\label{pF_true_d}
\begin{split}
p_F&=\mathbb{P}\left(\sum_{i=1}^dx_i>y^*\right)=\mathbb{P}\left(\frac{\sum_{i=1}^dx_i}{\sqrt{d}}>\frac{y^*}{\sqrt{d}}\right)\\
&=1-\Phi\left(\frac{y^*}{\sqrt{d}}\right).
\end{split}
\end{equation}
This expression will be used as a check on the SS estimate.

\begin{figure}[t]\centering
\includegraphics[angle=0,scale=0.47]{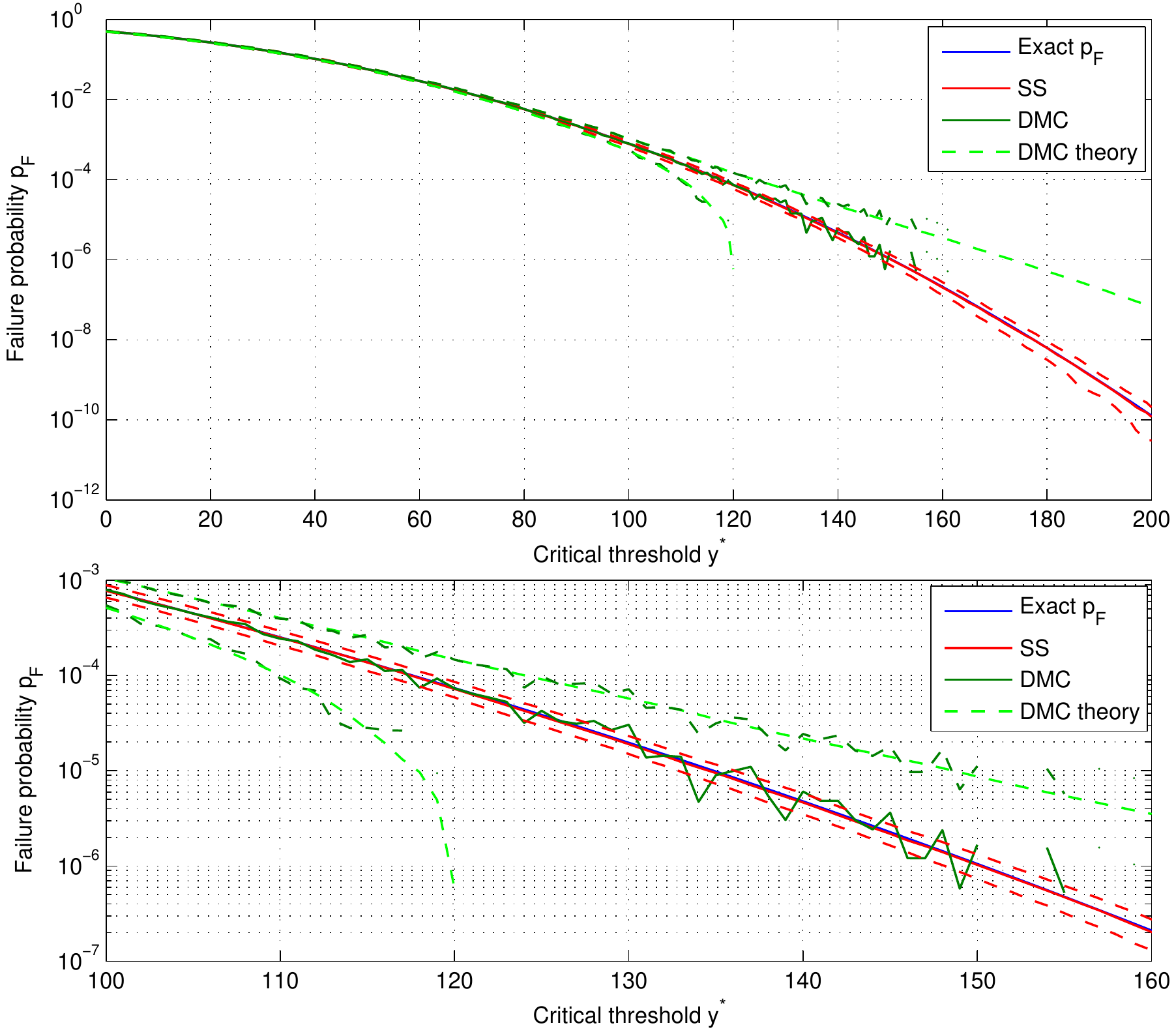}
\caption{ Critical threshold $y^*$ versus the failure probability $p_F$. [Example 6.2].} \label{FigA}
\end{figure}

First, consider the following range of values for the critical threshold, $y^*\in[0,200]$.
Fig.~\ref{FigA} plots $y^*$ versus $p_F$.
The solid red curve corresponds to the sample mean of the SS estimates $\hat{p}_F^{\mbox{\tiny{SS}}}$
which is based on $100$ independent runs of Subset Simulation. The two dashed red curves correspond to the sample mean $\pm$
one sample standard deviation. The SS parameters were set as follows:  the level probability $p=0.1$, the proposal PDFs $q_k(x|x_k)=\phi(x-x_k)$,
and the sample size $n=3\times10^3$ per each level.
The solid blue curve (which almost coincides with the solid red curve) corresponds to
the true values of $p_F$ computed from (\ref{pF_true_d}). The dark green curves correspond to Direct Monte Carlo:
the solid curve is the sample mean (based on 100 independent runs) of the DMC estimates $\hat{p}_F^{\mbox{\tiny{DMC}}}$ (\ref{pFDMC}), and
the two dashed curves are the sample mean $\pm$ one sample standard deviation. The total number of samples $N$ used in DMC equals
to the average (based on 100 runs) total number of samples used in SS. Finally, the dashed light green curves
show the theoretical performance of Direct Monte Carlo, namely, they correspond to the true value of $p_F$ (\ref{pF_true_d}) $\pm$
one theoretical standard deviation obtained from (\ref{V of DMC}). The bottom panel of Fig.~\ref{FigA} shows the zoomed in
region that corresponds to the values $y^*\in[100,160]$ of the critical threshold. Note that for relatively large values
of the failure probability, $p_F<10^{-3}$, both DMC and SS produce accurate estimates of $p_F$. For smaller values however, $p_F<10^{-5}$,
the DMC estimate starts to degenerate, while SS still accurately estimates $p_F$.
This can be seen especially well in the bottom panel of the figure.

\begin{figure}[t]\centering
\includegraphics[angle=0,scale=0.45]{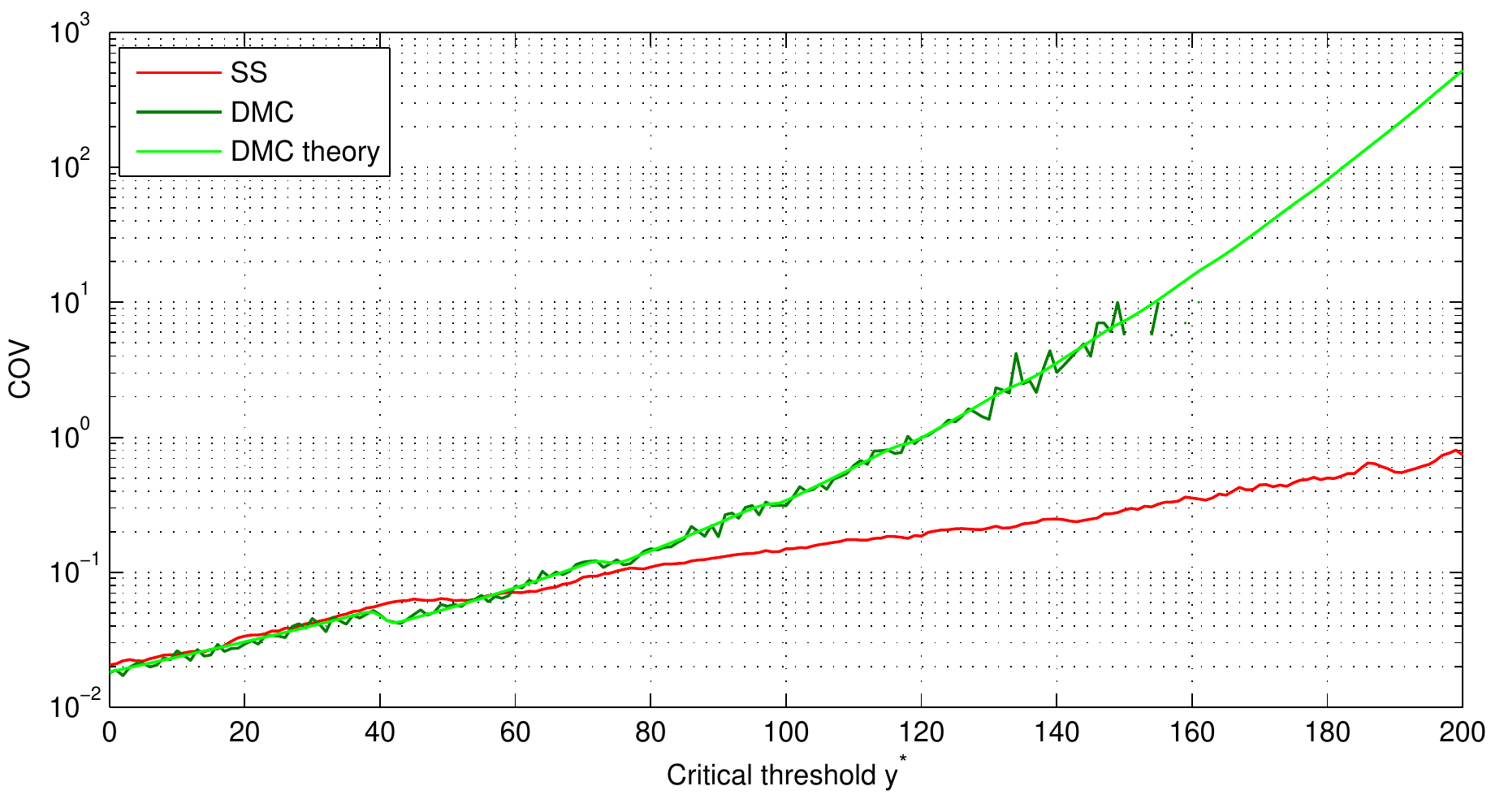}
\caption{Critical threshold $y^*$ versus the c.o.v. [Example 6.2].} \label{FigB}
\end{figure}

The performances of Subset Simulation and Direct Monte Carlo can be also compared in terms of the coefficient of variation of the estimates
$\hat{p}_F^{\mbox{\tiny{SS}}}$ and $\hat{p}_F^{\mbox{\tiny{DMC}}}$. This comparison is shown in Fig.~\ref{FigB}.
The red and dark green curves represent the sample c.o.v. for SS and DMC, respectively.
The light green curve is the theoretical c.o.v. of $\hat{p}_F^{\mbox{\tiny{DMC}}}$ given
by (\ref{cov_of_DMC}). When the critical threshold is relatively small $y^*<60$, the performances of SS and DMC are comparable.
As $y^*$ gets large, the c.o.v. of $\hat{p}_F^{\mbox{\tiny{DMC}}}$ starts to grow much faster than that of $\hat{p}_F^{\mbox{\tiny{SS}}}$.
In other words, SS starts to outperform DMC, and the larger $y^*$, i.e. the more rare the failure event, the more significant the outperformance is.

\begin{figure}[t]\centering
\includegraphics[angle=0,scale=0.45]{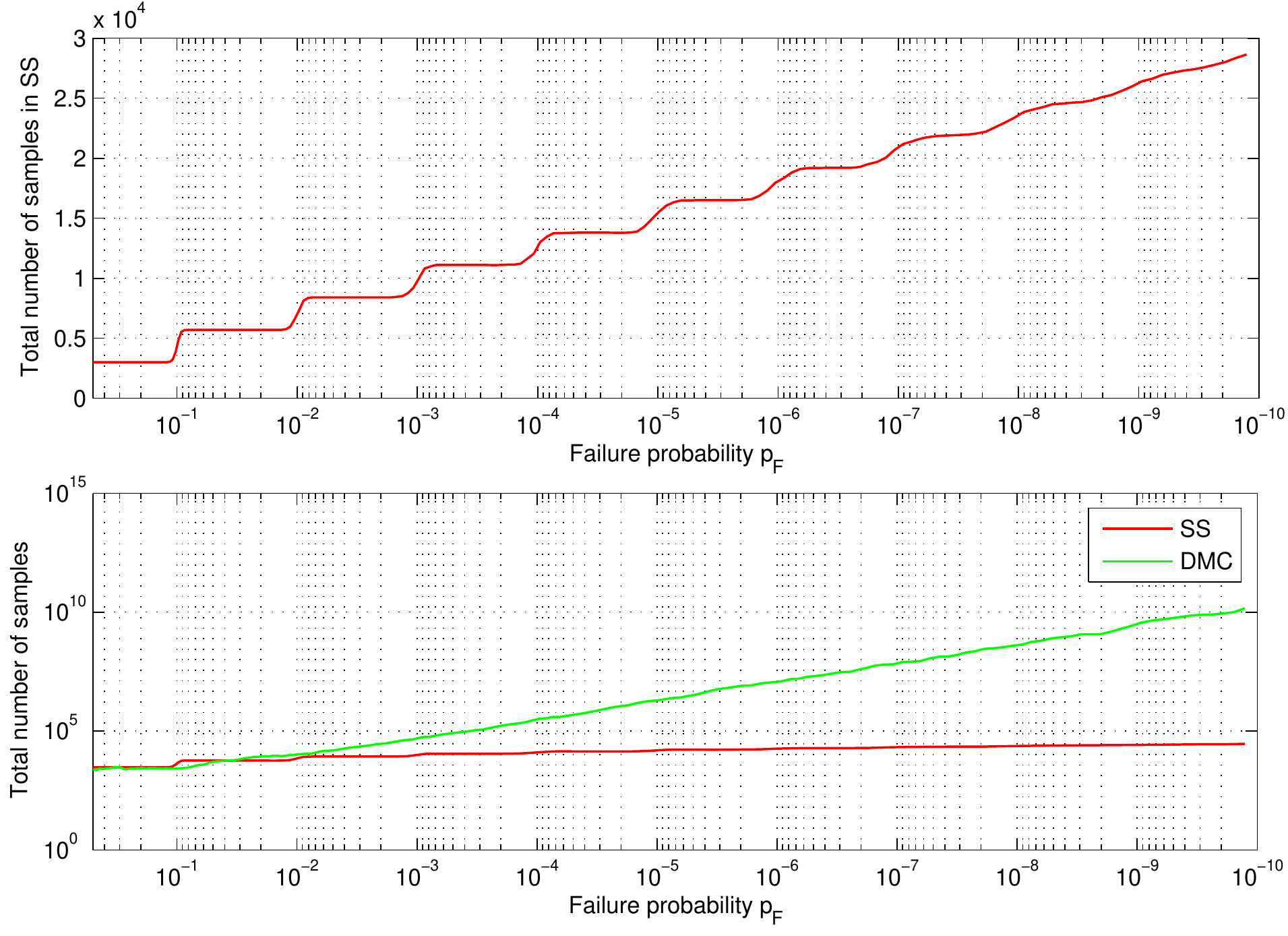}
\caption{ Failure probability versus the total number of samples. [Example 6.2].} \label{FigC}
\end{figure}

The average total number of samples used in Subset Simulation versus the corresponding values of failure probability
is shown in the top panel of Fig.~\ref{FigC}.
\begin{figure}[t]\centering
	\includegraphics[angle=0,scale=0.45]{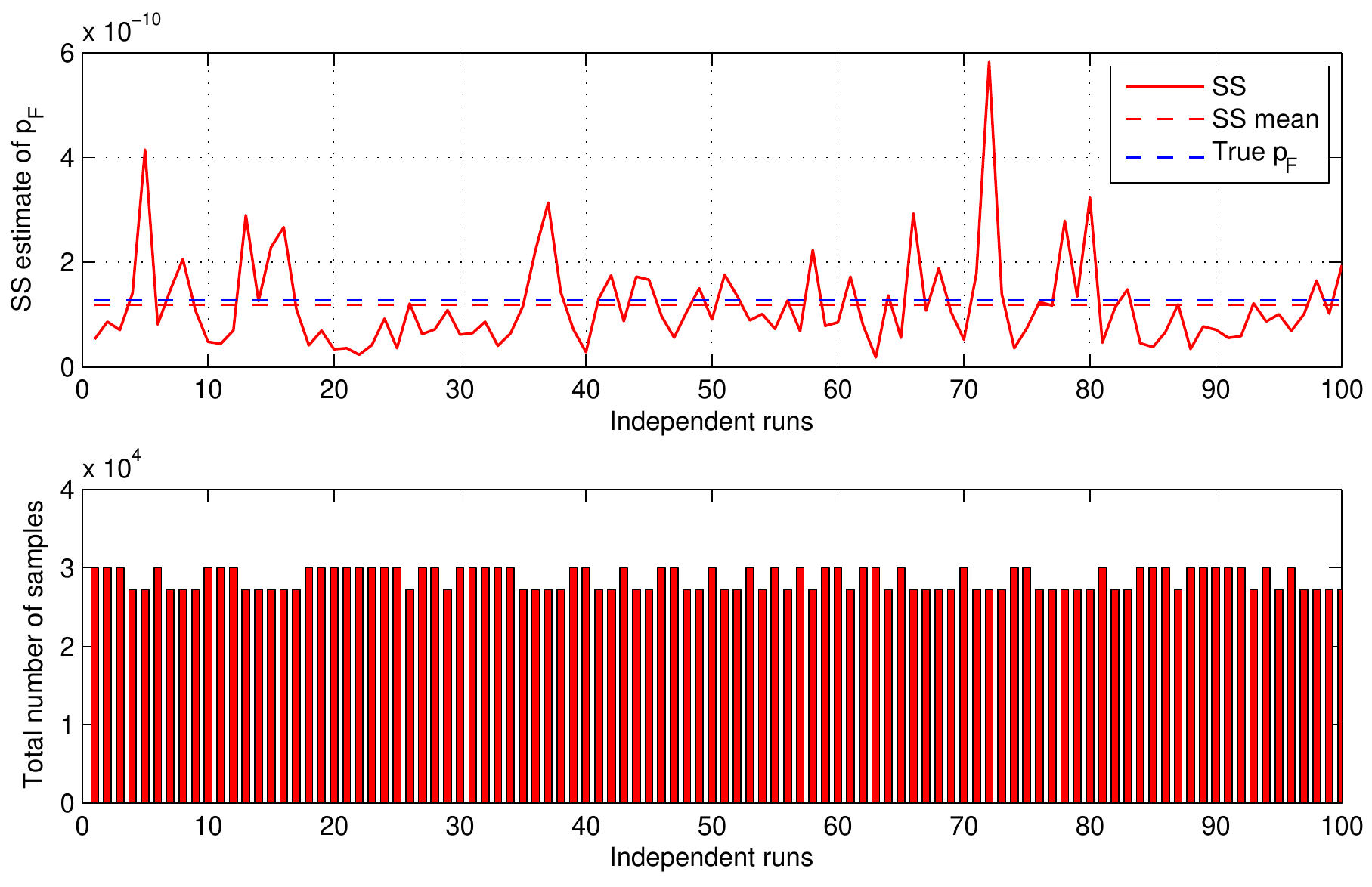}
	\caption{\footnotesize Performance of Subset Simulation for 100 independent runs. The critical threshold is $y^*=200$, the corresponding
		true value of the failure probability is $p_F=1.27\times10^{-10}$. [Example 6.2].} \label{FigD}
	\includegraphics[angle=0,scale=0.4]{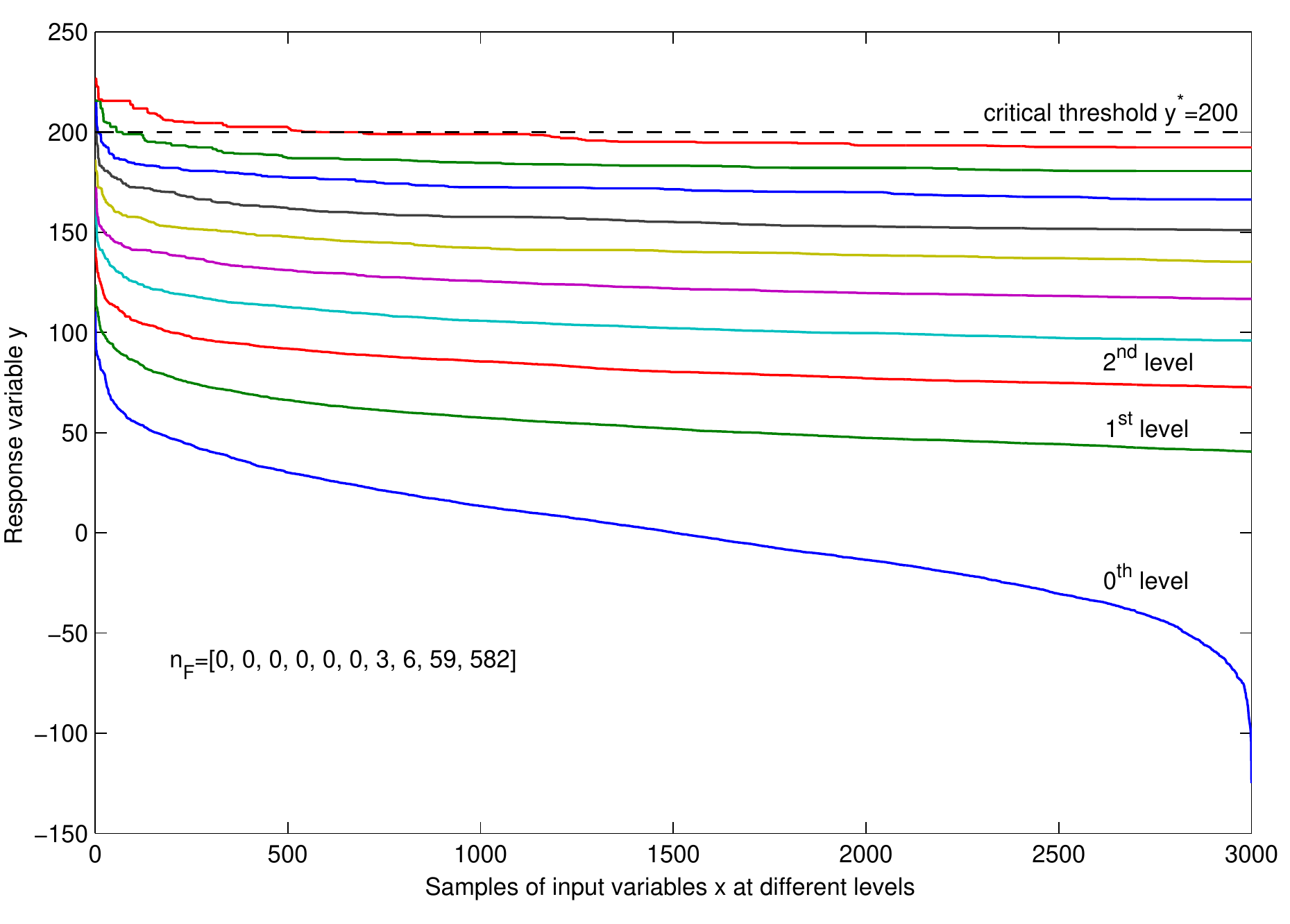}
	\caption{\footnotesize System responses $y_l^{(1)}\geq\ldots\geq y_l^{(n)}$, $n=3\times10^3$, for all levels, $l=0,\ldots,L=9$, for a fixed
		simulation run. [Example 6.2].} \label{FigE}
\end{figure} 
The staircase nature of the plot is due to the fact that every time when $p_F$ crosses the
value $p^{k}$ by decreasing from $p^{k}+\epsilon$ to $p^{k}-\epsilon$, an additional conditional level is required. In this example, $p=0.1$ is used,
that why the jumps occur at $p_F=10^{-k}$, $k=1,2,\ldots$.
The jumps are more pronounced for larger values of $p_F$, where the SS estimate is more
accurate. For smaller values of $p_F$, where the SS estimate is less accurate, the jumps are more smoothed out by averaging over independent runs.

In Fig.~\ref{FigB}, where the c.o.v's of SS and DMC are compared, the total numbers of samples (computational efforts)
used in the two methods are the same. The natural question is then the following: by how much should the total number of samples $N$
used in DMC be increased to achieve the same c.o.v as in SS (so that the green curve in Fig.~\ref{FigB} coincides
with the red curve)? The answer is given in the bottom panel of Fig.~\ref{FigC}. For example, if $p_F=10^{-10}$, then $N=10^{10}$,
while the computational effort of SS is less than $10^5$ samples.

Simulation results presented in Figures~\ref{FigA},\ref{FigB}, and \ref{FigC}
clearly indicate that (a) Subset Simulation produces a relatively accurate estimate of the failure probability,
and (b) Subset Simulation drastically outperforms Direct Monte Carlo when estimating probabilities of rare events.

Let us now focus on a specific value of the critical threshold, $y^*=200$, which corresponds to a very rare failure event (\ref{Flineard}) with
probability $p_F=1.27\times10^{-10}$. Fig.~\ref{FigD} demonstrates the performance of Subset Simulation for 100 independent runs.
The top panel shows the obtained SS estimate $\hat{p}_F^{\mbox{\tiny{SS}}}$ for each run. Although $\hat{p}_F^{\mbox{\tiny{SS}}}$
varies significantly (its c.o.v. is $\delta(\hat{p}_F^{\mbox{\tiny{SS}}})=0.74$), its mean value
$\overline{\hat{p}_F^{\mbox{\tiny{SS}}}}=1.18\times10^{-10}$ (dashed red line)
is close to the true value of the failure probability (dashed blue line).
The bottom panel shows the total number of samples used in SS in each run. It is needless to say that the DMC estimate based on
$N\sim 3\times10^4$ samples would almost certainly be zero.

\begin{figure}[t]\centering
\includegraphics[angle=0,scale=0.45]{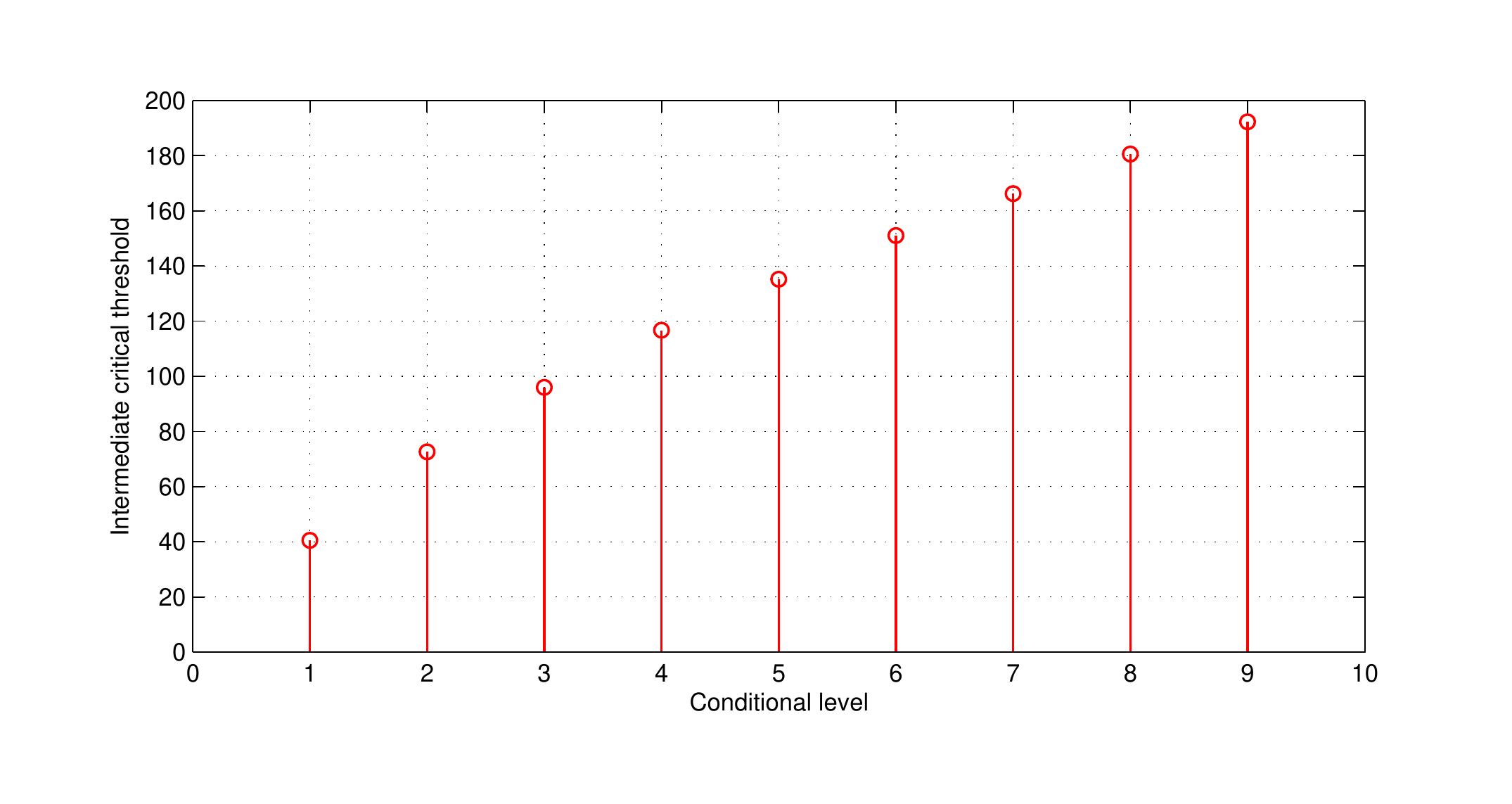}
\caption{\footnotesize Intermediate critical thresholds $y^*_1,\ldots,y^*_L$, $L=9$, at different conditional levels
in a fixed simulation run. [Example 6.2].} \label{FigF}
\end{figure}

Fig.~\ref{FigE} shows the system responses $y_l^{(1)}\geq\ldots\geq y_l^{(n)}$, $n=3\times10^3$, for all levels, $l=0,\ldots,L=9$, for a fixed
simulation run. As expected, for the first few levels (6 levels in this case), the number of failure samples $n_F(l)$, i.e. samples $x_l^{(i)}$ with
$y_l^{(i)}=g(x_l^{(i)})>y^*$, is zero. As Subset Simulation starts pushing the samples towards the failure domain, $n_F(l)$ starts increasing with $n_F(6)=3$,
$n_F(7)=6$, $n_F(8)=59$, and, finally, $n_F(9)=582$, after which the algorithm stopped since $n_F(9)/n=0.194$ which is large than $p=0.1$.
Finally, Fig.~\ref{FigF} plots the intermediate (relaxed) critical thresholds $y^*_1,\ldots,y^*_L$ at different levels obtained in a fixed simulation run.

\section{MATLAB code}\label{Appendix}
This section contains the MATLAB code for the examples considered in Section \ref{Examples}.
For educational purposes, the code was written as readable as possible with numerous comments.
As a result of this approach, the efficiency of the code was unavoidably scarified.
This code is also available online at \url{http://arxiv.org/}.
\begin{widetext} 
\begin{lstlisting}
% Subset Simulation for Liner Reliability Problem
% Performance function: g(x)=x1+...+xd
% Input variables x1,...,xd are i.i.d. N(0,1)
% Written by K.M. Zuev, Institute of Risk & Uncertainty, Uni of Liverpool
%------------------------------------------------------------------------
clear;
d=1000;                      % dimension of the input space
YF=200;                      % critical threshold (failure <=> g(x)>YF)
pF=1-normcdf(YF/sqrt(d));    % true value of the failure probability
n=3000;                      % number of samples per level
p=0.1;                       % level probability
nc=n*p;                      % number of Markov chains
ns=(1-p)/p;                  % number of states in each chain
%------------------------------------------------------------------------
L=0;                         % current (unconditional) level
x=randn(d,n);                % Monte Carlo samples
nF=0;                        % number of failure samples
for i=1:n
    y(i)=sum(x(:,i));        % system response y=g(x)
    if y(i)>YF               % y(i)>YF <=> x(:,i) is a failure sample
        nF=nF+1;
    end
end
while nF(L+1)/n<p            % stopping criterion
    L=L+1;                   % next conditional lelvel is needed
    [y(L,:),ind]=sort(y(L,:),'descend'); % renumbered responses
    x(:,:,L)=x(:,ind(:),L);              % renumbered samples
    Y(L)=(y(L,nc)+y(L,nc+1))/2;          % L^th intermediate threshold
    z(:,:,1)=x(:,1:nc,L);                % Markov chain "seeds"
    %--------------------------------------------------------------------
    % Modified Metropolis algorithm for sampling from pi(x|F_L)
    for j=1:nc
        for m=1:ns
            % Step 1:
            for k=1:d
                a=z(k,j,m)+randn;                       % Step 1(a)
                r=min(1,normpdf(a)/normpdf(z(k,j,m)));  % Step 1(b)
                % Step 1(c):
                if rand<r
                    q(k)=a;
                else
                    q(k)=z(k,j,m);
                end
            end
            % Step 2:
            if sum(q)>Y(L) % q belongs to F_L
                z(:,j,m+1)=q;
            else
                z(:,j,m+1)=z(:,j,m);
            end
        end
    end
    for j=1:nc
        for m=1:ns+1
            x(:,(j-1)*(ns+1)+m,L+1)=z(:,j,m); % samples from pi(x|F_L)
        end
    end
    clear z;
    %--------------------------------------------------------------------
    nF(L+1)=0;
    for i=1:n
        y(L+1,i)=sum(x(:,i,L+1)); % system response y=g(x)
        if y(L+1,i)>YF            % then x(:,i,L+1) is a failure sample
            nF(L+1)=nF(L+1)+1;    % number of failure samples at level L+1
        end
    end
end
pF_SS=p^(L)*nF(L+1)/n;            % SS estimate
N=n+n*(1-p)*(L);                  % total number of samples
%------------------------------------------------------------------------
\end{lstlisting}
\end{widetext}

\section{Summary}\label{Conclusions}

In this paper, a detailed exposition  of Subset Simulation, an advanced stochastic simulation method for estimation
of small probabilities of rare events, is provided at introductory level. A simple step-by-step derivation of Subset Simulation is given,
and important implementation details are discussed. The method is illustrated with a few intuitive examples.

After the original paper \cite{AuBeck} was published, various modifications of SS were proposed: SS with
splitting \cite{ChingAuBeck}, Hybrid SS \cite{ChingBeckAu}, and
Two-Stage SS \cite{KatafygiotisCheung}, to name but a few. It is important to
highlight, however, that none of these modifications offer a drastic
improvement over the original algorithm. A Bayesian analog of SS was developed in \cite{BSS}.
For further reading on Subset Simulation and its applications, a fundamental and very accessible monograph \cite{SSbook} is strongly recommended,
where the method is presented from the CCDF (complementary cumulative distribution function) perspective and where the error estimation
is discussed in detail.

Also, it is important to emphasize  that Subset Simulation provides an
efficient solution for general reliability problems without using any
specific information about the dynamic system other than an
input–output model. This independence of a system's inherent
properties makes Subset Simulation potentially useful for
applications in different areas of science and engineering.


As a final remark, it is a pleasure to thank Professor Siu-Kui Au whose comments on the first draft of the paper were very helpful,
Professor James Beck, who generously shared his knowledge of and experience with Subset Simulation, and Professor
Francis Bonahon for general support and for creating a nice atmosphere at the Department of Mathematics of the University of Southern California, where the author
started this work.

\end{document}